\documentclass[11pt,a4paper]{article}
\usepackage{jheppub}
\pdfoutput=1

\usepackage[utf8]{inputenc} 
\usepackage[T1]{fontenc} 
\usepackage{lmodern}
\usepackage[british]{babel}

\usepackage{microtype} 
\makeatletter
\g@addto@macro\@floatboxreset\centering
\makeatother

\usepackage{xfrac}
\numberwithin{equation}{section} 

\usepackage[capitalise]{cleveref} 

\usepackage{booktabs} 
\usepackage{array} 
\newcolumntype{L}{>{$}l<{$}} 

\usepackage[italic]{hepnames} 
\renewcommand{\PBs}{\HepParticle{\PB}{\Pqs}{}\xspace}
\renewcommand{\APBs}{\HepAntiParticle{\PB}{\Pqs}{}\xspace}

\renewcommand{\PZprime}{\ensuremath{Z'}\xspace}
\usepackage{braket} 
\usepackage{slashed} 
\usepackage{siunitx} 
\DeclareSIUnit\fb{\femto\barn}
\DeclareSIUnit\invps{\ps^{-1}}
\usepackage{suffix} 

\newcommand{\Lagrangian}{\ensuremath{\mathcal{L}}\xspace}
\newcommand{\DeltaMs}{\Delta M_s}
\newcommand{\DeltaMd}{\Delta M_d}
\newcommand{\Acpmix}{A^\text{CP}_\text{mix}}
\newcommand{\RKRKstar}{R_{K^{(*)}}}
\newcommand{\RK}{R_K}
\newcommand{\RKstar}{R_{K^*}}
\newcommand{\V}[1]{\ensuremath{V_{#1}^{}}} 
\WithSuffix\newcommand\V*[1]{\ensuremath{V_{#1}^*}}

\allowdisplaybreaks

\title{
\texorpdfstring{$\Delta M_s$}{DeltaMs} theory precision confronts flavour anomalies
}

\author[a]{Luca Di Luzio}
\author[b]{Matthew Kirk}
\author[c]{Alexander Lenz}
\author[d]{Thomas Rauh}

\affiliation[a]{Dipartimento di Fisica ``E.~Fermi'', Universit\`a di Pisa and INFN Sezione di Pisa, \\
Largo Bruno Pontecorvo 3, I-56127 Pisa, Italy}
\affiliation[b]{Dipartimento di Fisica, Universit\`a di Roma ``La Sapienza'', 
and INFN Sezione di Roma, Piazzale Aldo Moro 2, 00185 Roma, Italy}
\affiliation[c]{Institute for Particle Physics Phenomenology, Durham University, \\
DH1 3LE Durham, United Kingdom}
\affiliation[d]{Albert Einstein Center for Fundamental Physics, Institute for Theoretical Physics, University of Bern, Sidlerstrasse 5, CH-3012 Bern, Switzerland}
\emailAdd{luca.diluzio@pi.infn.it}
\emailAdd{matthew.kirk@roma1.infn.it}
\emailAdd{alexander.lenz@durham.ac.uk}
\emailAdd{rauh@itp.unibe.ch}

\preprint{IPPP/19/70}
\arxivnumber{1909.11087}

\abstract{
Based on recent HQET sum rule and lattice calculations we present updated 
Standard Model predictions for the mass differences of neutral $B$ mesons: 
$\Delta M_s^{\rm SM} = \left(18.4^{+0.7}_{-1.2} \right) \mbox{ps}^{-1}$
and
$\Delta M_d^{\rm SM} = \left(0.533^{+0.022}_{-0.036} \right) \mbox{ps}^{-1}$ 
and study their impact on new physics models that address the present hints of 
anomalous data in $b \to s \ell \ell$ transitions. We also examine future prospects 
of further reducing the theory uncertainties and discuss the implications of a 
2025 scenario with 
$\Delta M_s^\text{SM 2025} = \SI{18.4 \pm 0.5}{\invps}$. 
In particular, the latter yields upper bounds $M_{Z'}\lesssim \SI{9}{\TeV}$ and 
$M_{S_3}\lesssim \SI{30}{\TeV}$ for the minimal $Z'$ and $S_3$ lepto-quark explanations 
of the $b \to s \ell \ell$ anomalies, respectively. 
}

\begin{document}

\maketitle
\section{Introduction\label{sec:intro}}

The mixing of neutral $B$ mesons provides an important test of our understanding of the Standard Model (SM). 
It gives direct access to the poorly known CKM elements $V_{td}$,  $V_{ts}$ and $V_{tb}$,  and it is an
excellent probe for physics beyond the standard model (BSM), 
see e.g.~\cite{Artuso:2015swg}.
\\
By now the experimental values of the mass differences of the neutral $B$ mesons
are known very precisely \cite{Amhis:2016xyh} (based on the measurements in
\cite{Prentice:1987ap,Abulencia:2006ze,Aaij:2011qx,Aaij:2012mu,Aaij:2012nt,Aaij:2013mpa,Aaij:2013gja,Aaij:2014zsa})
\begin{align}
  \DeltaMd^\text{exp} &= \SI{0.5064 \pm 0.0019}{\invps} \label{eq:DeltaMdExp} \, ,
  \\
  \DeltaMs^\text{exp} &= \SI{17.757 \pm 0.021}{\invps} \label{eq:DeltaMsExp} \, .
\end{align}
The theoretical determination of the mass differences is limited by our understanding of 
non-perturbative matrix elements of dimension six operators. The matrix elements can be 
determined with lattice simulations or sum rules. Both approaches utilize a three-point 
correlator of two interpolating currents for the $B_q$ and $\bar{B}_q$ mesons and the 
dimension six operator, which is significantly more complicated than the determination of 
decay constants where one considers a two-point correlator of two currents. 
Therefore, for a long time (see e.g.~the FLAG 2013 review \cite{Aoki:2013ldr}) the precision 
of the matrix elements was limited to more than 10\%. The consequence were SM predictions 
of $\Delta M_q$ \cite{Artuso:2015swg} close to the experimental numbers, but with much 
higher uncertainties: 
\begin{align}
\DeltaMd^\text{FLAG 2013} &= \SI{0.528 \pm 0.078}{\invps} = \left(1.04\pm0.15\right) \DeltaMd^\text{exp} \label{eq:DeltaMdFLAG2013} \, ,
\\
\DeltaMs^\text{FLAG 2013} &= \SI{18.3 \pm 2.7}{\invps} = \left(1.03\pm0.15\right) \DeltaMs^\text{exp} \label{eq:DeltaMsFLAG2013} \, .
\end{align}
In 2016 \cite{Bazavov:2016nty} the FNAL/MILC collaboration presented a new evaluation of 
the non-perturbative $B$ mixing matrix elements yielding larger values than previously 
obtained. They achieved relative errors of 8.5\% and 6.4\% for the matrix elements relevant 
for $B_d$ and $B_s$ mixing in the SM, respectively, thus pushing the uncertainty of simulations 
with 2$+1$ dynamical flavours below the 10\% benchmark for the first time. The FNAL/MILC results 
dominate the current FLAG average \cite{Aoki:2019cca} and we
get the following predictions\footnote{In \cite{Bazavov:2016nty} FNAL/MILC quote as predictions 
  $\DeltaMd^\text{FNAL/MILC 2016} = \SI{0.630 \pm 0.069}{\invps} $ and 
  $\DeltaMs^\text{FNAL/MILC 2016} = \SI{19.6 \pm 1.6}{\invps}$.
  The small difference with respect to our quoted values stems from a different 
  treatment of CKM inputs.} 
\begin{align}
\DeltaMd^\text{FLAG 2019} &= \left(0.582_{-0.056}^{+0.049}\right)\text{ps}^{-1} = \left(1.15_{-0.11}^{+0.10}\right) \DeltaMd^\text{exp} \label{eq:DeltaMdFLAG2019} \, ,
\\
\DeltaMs^\text{FLAG 2019} &= \left(20.1_{-1.6}^{+1.2}\right)\text{ps}^{-1} = \left(1.13_{-0.09}^{+0.07}\right) \DeltaMs^\text{exp} \label{eq:DeltaMsFLAG2019} \, ,
\end{align}
which, when combined, are about two standard deviations above the experimental numbers. 
The values in Eq.~(\ref{eq:DeltaMdFLAG2019}) and Eq.~(\ref{eq:DeltaMsFLAG2019})
for the mass differences could be the starting point of new anomalies arising
in the $b$-sector, see e.g.~\cite{Blanke:2016bhf,Buras:2016dxz,Bobeth:2017xry,Blanke:2018cya,Hu:2019heu}.
In addition $\Delta M_s$ poses very severe constraints on certain BSM models -- in particular 
\PZprime models -- that address the observed anomalies
in $b \to s \ell \ell$ transitions, see e.g.~\cite{DiLuzio:2017fdq}.

\begin{figure}
\includegraphics[width=0.65\textwidth]{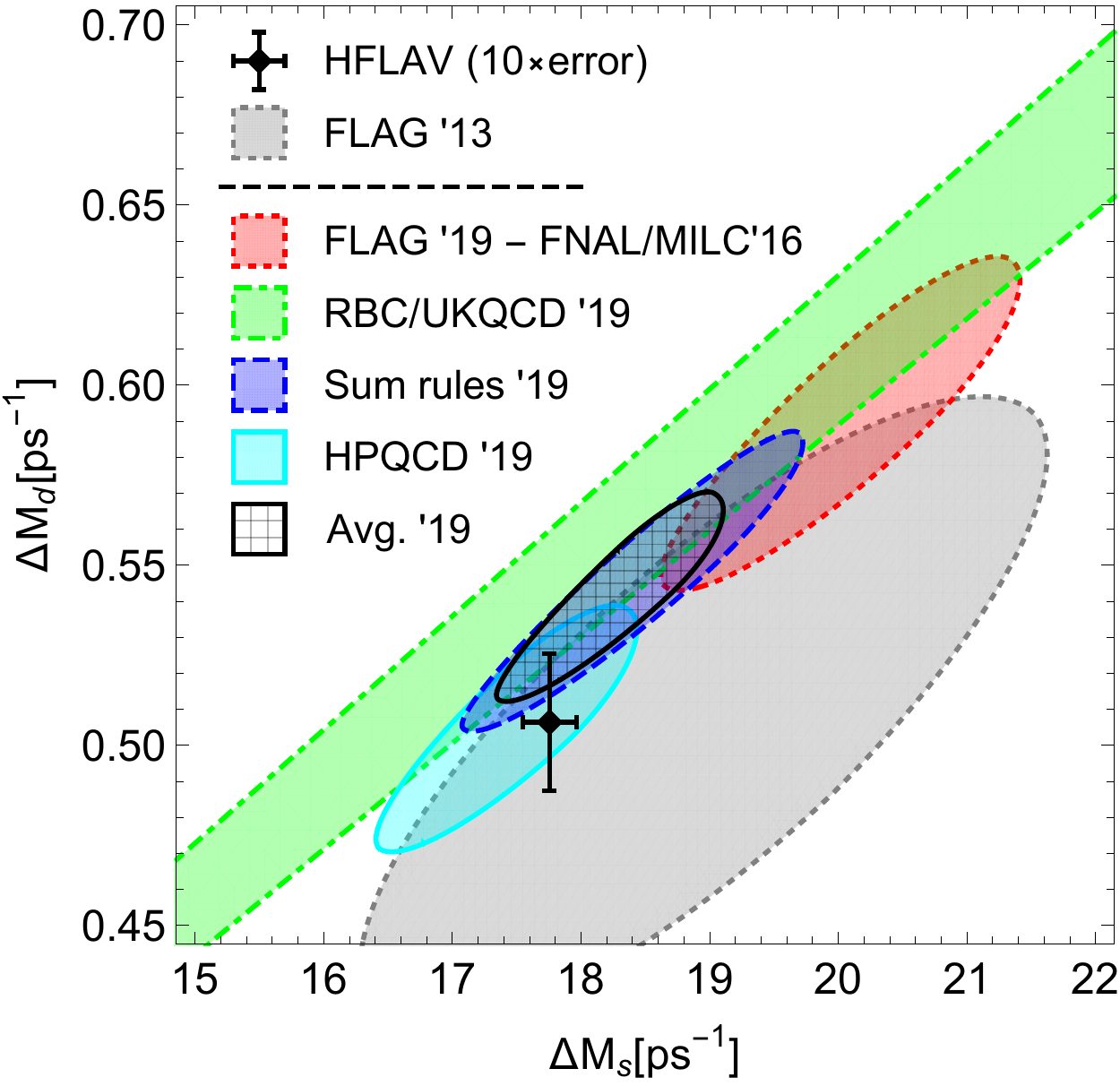}
\caption{Predictions for $\Delta M_s$ and $\Delta M_d$. See Section~\ref{sec:SM} for details.}
\label{fig:DeltaMsVsDeltaMd}
\end{figure}

To date several new determinations of the matrix elements with a precision below 10\% 
have appeared: first from sum rules~\cite{Grozin:2016uqy,Grozin:2017uto,Grozin:2018wtg,Kirk:2017juj,King:2019lal} 
and very recently from the HPQCD collaboration \cite{Dowdall:2019bea}. The SU(3) 
breaking ratios were also evaluated by the RBC/UKQCD collaboration \cite{Boyle:2018knm}. 
We compare the various predictions in Fig.~\ref{fig:DeltaMsVsDeltaMd} where also the 
results using a weighted average of the values of 
\cite{Aoki:2019cca,Kirk:2017juj,King:2019lal,Dowdall:2019bea,Boyle:2018knm} for the matrix 
elements are shown. The weighted average gives an impressive leap of the precision to 
3.6\% and 3.1\% for the matrix elements for $B_d$ and $B_s$ mixing. We observe excellent 
agreement between the sum rule prediction and the weighted average and given that lattice 
simulations and sum rules have very different systematic uncertainties this is a strong 
confirmation of both methods. We discuss the status of the non-perturbative input in 
more detail in Section~\ref{sec:SM}. 
With respect to Eq.~\eqref{eq:DeltaMdFLAG2019} and Eq.~\eqref{eq:DeltaMsFLAG2019} 
the weighted averages 
\begin{align}
\DeltaMd^\text{Average 2019} &= \left(0.533_{-0.036}^{+0.022}\right)\text{ps}^{-1} = \left(1.05_{-0.07}^{+0.04}\right) \DeltaMd^\text{exp} \label{eq:DeltaMdAVG2019} \, ,
\\
\DeltaMs^\text{Average 2019} &= \left(18.4_{-1.2}^{+0.7}\right)\text{ps}^{-1} = \left(1.04_{-0.07}^{+0.04}\right) \DeltaMs^\text{exp} \label{eq:DeltaMsAVG2019} \, ,
\end{align}
show better agreement with experiment and a reduction of the total errors by about 40\% 
to the point where the hadronic and parametric CKM uncertainties are of the same size. 

We study the mass difference within the SM in more detail in Section~\ref{sec:SM} and 
give our estimate for the accuracy of the matrix elements in about five years time. 
Together with the extrapolation for the precision of the CKM elements presented in 
\cite{Kou:2018nap} we expect 
\begin{align}
\DeltaMs^\text{Future 2025} &= \left(18.4\pm0.5\right)\text{ps}^{-1} = \left(1.04\pm0.03\right) \DeltaMs^\text{exp} \label{eq:DeltaMsFuture} \, .
\end{align}
for a future ($\sim$~2025) scenario when the current $b\to s\ell\ell$ anomalies should be 
established at the level of about 10 standard deviations if the central values remain
the same~\cite{Bediaga:2018lhg}. In Section~\ref{BSM} we investigate the implications 
of $B_s$ mixing on the $b \to s \ell \ell$ anomalies in the FLAG~'19, Average~'19 and 
Future~'25 scenarios. First, we assume minimal $Z'$ and lepto-quark (LQ) scenarios with 
only the couplings required to address the anomalies. Then, we discuss the viability of 
model-building ideas beyond the minimal $Z'$ scenario that might reduce the theory value 
for $\DeltaMs$ and thus improve the agreement with experiment. 
Finally, we conclude in Section \ref{conclusions}.

\section{\texorpdfstring{\(\DeltaMs\)}{DeltaMs} in the Standard Model\label{sec:SM}}

\begin{figure}
\includegraphics[width=\textwidth]{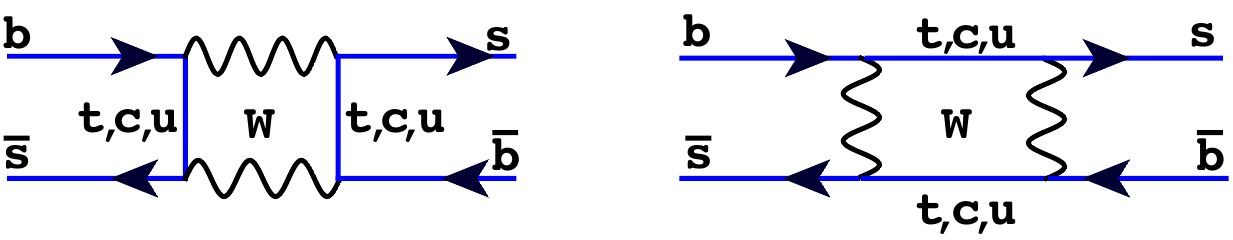}
\caption{Box diagrams that generate the \PBs-\APBs mixing process in the SM.
  $M_{12}^s$ is generated when the internal particles are off-shell.}
\label{fig:Bsmixing_box_SM}
\end{figure}

In the SM, \PBs mixing is generated by the diagrams shown in \cref{fig:Bsmixing_box_SM}.
The observable of interest in this work is the mass difference of the two mass eigenstates:
\begin{equation}
\DeltaMs \equiv M^s_H - M^s_L = 2 \left| M_{12}^s \right| \, .
\end{equation}
The SM calculation (see e.g.\ \cite{Artuso:2015swg} for a review) gives the following result for $M_{12}^s$
\begin{equation}
\label{eq:M12SM}
M_{12}^{s,\,\text{SM}} = \frac{G_F^2}{12 \pi^2} \lambda_t^2 M_W^2 S_0(x_t) \hat{\eta}_B f_{\PBs}^2 M_{\PBs} B_1 \,,
\end{equation}
where $\lambda_t = \V{tb}\V*{ts}$, $S_0$ is the Inami-Lim function \cite{Inami:1980fz},
$x_t = \left( \bar m_t(\bar m_t) / M_W \right)^2$ and $\hat{\eta}_B \approx 0.84$ encodes
the 2-loop perturbative QCD corrections \cite{Buras:1990fn}.
In the SM, only a single four-quark $\Delta B = 2$ operator arises,
\begin{equation}
\label{eq:operator_basis_Q1}
Q_1 = \APqs \gamma^\mu(1-\gamma^5) \Pqb \, \,  \APqs \gamma_\mu(1-\gamma^5) \Pqb \,,
\end{equation}
whose hadronic matrix element is parameterised in terms of the meson decay constant
$f_{\PBs}$ and a bag parameter $B_1$
\begin{equation}
\braket{Q_1} = \braket{\PBs|Q_1|\APBs} \equiv \frac{8}{3} M_{\PBs}^2 f_{\PBs}^2 B_1 (\mu_b) \,.
\label{defB1}
\end{equation}
Note that we have used a form of \cref{eq:M12SM} (matching that in \cite{Artuso:2015swg})
where the bag parameter $B_1$ depends on the \Pqb quark scale $\mu_b \sim m_b$.
An alternative notation (commonly used by lattice groups, for example see FLAG \cite{Aoki:2019cca})
uses the pair $(\eta_B, \hat{B}_1)$ instead of $(\hat{\eta}_B, B_1(\mu_b))$, with $\hat{B}_1$
a renormalization group (RG) invariant quantity.
The two are related such that their product is equal, and explicitly we have
\begin{equation}
\hat{B}_1 = \alpha_s(\mu_b)^{-6/23} \left( 1 + \frac{\alpha_s(\mu_b)}{4\pi} \frac{5165}{3174} \right) B_1(\mu_b) = 1.519 \, B_1 (\mu_b) \,.
\end{equation}
The combination $f_{\PBs}^2 B_1$ is the least-well known parameter for \PBs mixing, and
the most important as the observable $\DeltaMs$ is directly proportional to it.
In the SM determination of the decay rate difference $\Delta \Gamma_q$ and for BSM contributions to
$\Delta M_q$ in addition to $Q_1$ four more operators arise:
\begin{align}
 Q_2 & = \bar{s}_i(1-\gamma^5)b_i\,\,\bar{s}_j(1-\gamma^5)b_j\,,\hspace{1cm}
 Q_3   = \bar{s}_i(1-\gamma^5)b_j\,\,\bar{s}_j(1-\gamma^5)b_i\,,\nonumber\\
 Q_4 & = \bar{s}_i(1-\gamma^5)b_i\,\,\bar{s}_j(1+\gamma^5)b_j\,,\hspace{1cm}
 Q_5   = \bar{s}_i(1-\gamma^5)b_j\,\,\bar{s}_j(1+\gamma^5)b_i\,, 
 \label{eq:operator_basis_Q25}
\end{align} 
which are typically parameterised as
\begin{align}
 \langle Q_2 \rangle & = f_{B_s}^2M_{B_s}^2\, \frac{-5M_{B_s}^2}{3(\bar{m}_b(\mu_b)+\bar{m}_s(\mu_b))^2} \, B_2(\mu_b)\,,\nonumber\\
 \langle Q_3 \rangle & = f_{B_s}^2M_{B_s}^2\, \frac{M_{B_s}^2}{3(\bar{m}_b(\mu_b)+\bar{m}_s(\mu_b))^2} \, B_3(\mu_b)\,,\nonumber\\
 \langle Q_4 \rangle & = f_{B_s}^2M_{B_s}^2\, \left[\frac{2M_{B_s}^2}{(\bar{m}_b(\mu_b)+\bar{m}_s(\mu_b))^2}+\frac{1}{3}\right] \, B_4(\mu_b)\,,\nonumber\\
 \langle Q_5 \rangle & = f_{B_s}^2M_{B_s}^2\, \left[\frac{2M_{B_s}^2}{3(\bar{m}_b(\mu_b)+\bar{m}_s(\mu_b))^2}+1\right] \, B_5(\mu_b)\,. 
 \label{eq:operator_basis_bags}
\end{align} 
The full basis of matrix elements has been determined with lattice 
simulations~\cite{Carrasco:2013zta,Bazavov:2016nty,Dowdall:2019bea} and 
sum rules~\cite{Kirk:2017juj,King:2019lal}. In \cite{Bazavov:2016nty} the combinations 
$f_{B_s}^2B_i$ were determined and combined with the 2016 PDG average for the decay constants 
to extract the Bag parameters. This implies that the latter suffer from bigger uncertainties 
because one cannot exploit partial cancellations of uncertainties which occur when $f_{B_s}$ 
is determined in the same simulation. All other works determine the Bag parameters $B_i(\mu_b)$ 
as their central results. This has the advantage that the values can easily be combined with 
updated results for the decay constant $f_{B_s}$ which has seen significant improvements in the 
last few years and we use the FLAG 2$+$1$+$1 average~\cite{Aoki:2019cca} of the 
results~\cite{Bussone:2016iua,Hughes:2017spc,Bazavov:2017lyh} below. 

In fact the sum rule method directly determines the Bag parameters and requires 
an independent result for the decay constant to obtain the matrix elements. In this approach 
one has to determine the perturbative 3-loop contribution to a correlator describing $B$ mixing.
Based on the master integrals presented in \cite{Grozin:2008nu} this calculation was first performed for
$B_d$ mixing by \cite{Grozin:2016uqy,Grozin:2017uto,Grozin:2018wtg} and confirmed in \cite{Kirk:2017juj}, where
this method was also extended to different mixing operators and to lifetime operators. In \cite{King:2019lal}
corrections due to a finite value of the strange quark mass were determined, yielding new predictions 
for the mass differences in better agreement with experiment than the FNAL/MILC values 
(\cref{eq:DeltaMdFLAG2019,eq:DeltaMsFLAG2019}) with similar precision. 

\begin{figure}
\includegraphics[width=0.75\textwidth]{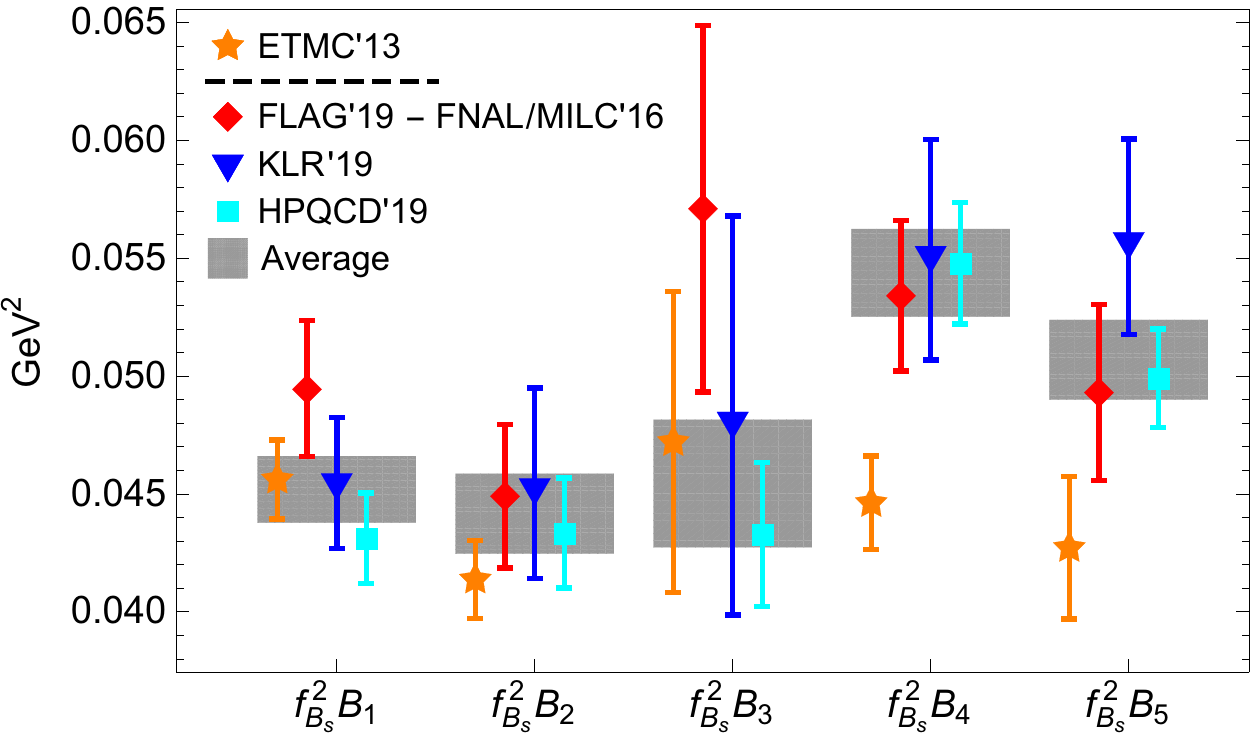}
\caption{Comparison of non-perturbative determinations of the matrix elements for $B_s$ mixing 
 at the scale $\mu_b=\bar{m}_b(\bar{m}_b)$.
 We show the ETMC~'13 values in orange, the FLAG 2+1 average~\cite{Aoki:2019cca} for the operator 
 $Q_1$ together with the FNAL/MILC~'16 results~\cite{Bazavov:2016nty} for the other operators in red, 
 the sum rule values~\cite{King:2019lal} in blue and the HPQCD~'19 results~\cite{Dowdall:2019bea} in 
 cyan. The gray rectangles indicate the weighted averages of the first three determinations.}
\label{fig:MatrixElementsComparison}
\end{figure}

Very recently an independent lattice study was performed by HPQCD \cite{Dowdall:2019bea} which 
does not confirm the large FNAL/MILC predictions for the mass differences. We present a comparison 
of the combinations $f_{B_s}^2B_i$ in Fig.~\ref{fig:MatrixElementsComparison}. Besides the individual 
results \cite{Carrasco:2013zta,Aoki:2019cca,Bazavov:2016nty,King:2019lal,Dowdall:2019bea} we show the 
weighted averages thereof where we have excluded \cite{Carrasco:2013zta} which only uses two dynamical 
flavours. For $f_{B_s}^2B_1$, which determines the SM value of the mass difference, 
the FLAG~'19 and HPQCD~'19 values do not overlap. However, the overall average shows perfect agreement 
with the KLR~'19 and ETMC~'13 results despite the very different systematic uncertainties of lattice 
simulation and sum rules. This makes us confident that the weighted average represents a reliable 
assessment of the current uncertainties. The different values of $f_{B_s}^2B_2$ agree well. 
The case of $f_{B_s}^2B_3$ is similar to that of the SM matrix element but the uncertainties are 
considerably larger. Finally for $f_{B_s}^2B_4$ and $f_{B_s}^2B_5$ we find a discrepancy between 
the results from ETMC~'13 and the remaining ones. This difference might result from the use of RI-MOM 
renormalisation scheme in \cite{Carrasco:2013zta}, see e.g.~the discussion in \cite{Garron:2016mva} 
for a similar problem in the $K$ sector. The other values are in good agreement with each other, with 
KLR~'19 yielding a somewhat larger result than the lattice simulations for $f_{B_s}^2B_5$. 
Our weighted averages at the scale $\mu_b=\bar{m}_b(\bar{m}_b)$ read 
\begin{align}
 f_{B_s}^2B_1(\mu_b) & = (0.0452\pm0.0014)\,\text{GeV}^2\,,\nonumber\\
 f_{B_s}^2B_2(\mu_b) & = (0.0441\pm0.0017)\,\text{GeV}^2\,,\nonumber\\
 f_{B_s}^2B_3(\mu_b) & = (0.0454\pm0.0027)\,\text{GeV}^2\,,\nonumber\\
 f_{B_s}^2B_4(\mu_b) & = (0.0544\pm0.0019)\,\text{GeV}^2\,,\nonumber\\
 f_{B_s}^2B_5(\mu_b) & = (0.0507\pm0.0017)\,\text{GeV}^2\,.
 \label{eq:BsAverageMatrixElements}
\end{align}
For convenience we also provide the weighted averages for $B_d$ mixing and the Bag parameters in \ref{sec:inputs}. 

\begin{figure}
\includegraphics[width=0.6\textwidth]{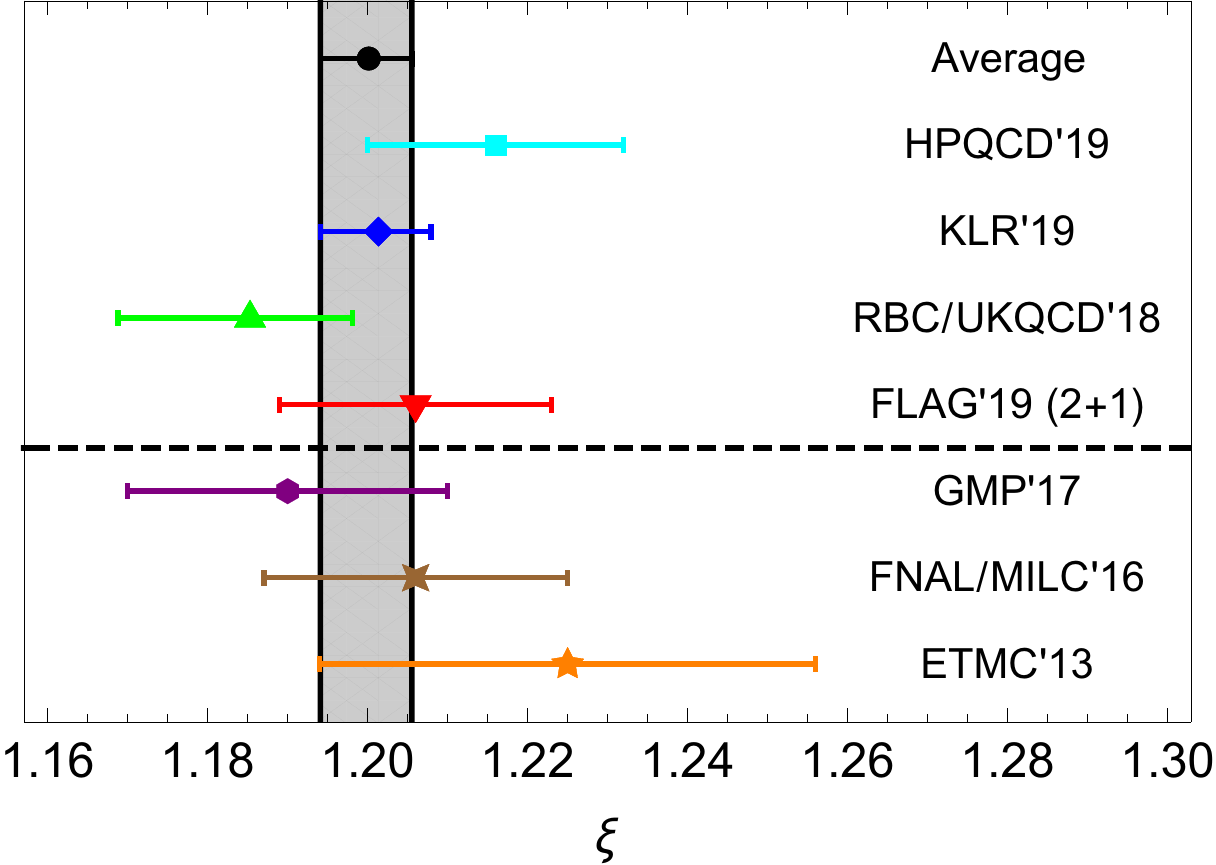}
\caption{Comparison of determinations of the SU(3)-flavour breaking ratio $\xi$. The gray band corresponds 
 to the weighted average of the results \cite{Aoki:2019cca,Boyle:2018knm,King:2019lal,Dowdall:2019bea}. 
 Below the dashed line we also show the values from \cite{Carrasco:2013zta,Bazavov:2016nty,Grozin:2017uto} 
 which are not included in the average.}
\label{fig:XiComparison}
\end{figure}

The ratio of the mass differences of $B_s$ and $B_d$ mesons benefits from the cancellation of many uncertainties 
and in the SM we get 
\begin{equation}
  \frac{\Delta M_d}{\Delta M_s} = \left| \frac{V_{td}^2}{V_{ts}^2} \right| \frac{M_{B_d}}{M_{B_s}} \frac{f_{B}^2 B_1^{B_d}}{f_{B_s}^2 B_1^{B_s}}
                                \equiv \left| \frac{V_{td}^2}{V_{ts}^2} \right| \frac{M_{B_d}}{M_{B_s}} \xi^{-2} \, .
\end{equation}
Different determinations \cite{Carrasco:2013zta,Bazavov:2016nty,Grozin:2017uto,Boyle:2018knm,Aoki:2019cca,King:2019lal,Dowdall:2019bea} 
of the ratio $\xi$ are compared in Fig.~\ref{fig:XiComparison} and we observe good agreement. The 
weighted average 
\begin{equation}
 \xi = 1.200_{-0.0060}^{+0.0054}
 \label{eq:xi_avg}
\end{equation}
of the values from \cite{Boyle:2018knm,Aoki:2019cca,King:2019lal,Dowdall:2019bea} is dominated 
by the sum rule result \cite{King:2019lal} which is about a factor of two more precise than the 
lattice determinations. Using the CKMFitter result 
\begin{equation}
 |V_{td}/V_{ts}| = 0.2088_{-0.0030}^{+0.0016}\,,
 \label{eq:VtdVts}
\end{equation}
the ratio of the mass differences becomes 
\begin{align}
 \label{eq:DeltaMdOverDeltaMsExp}
 \left(\frac{\Delta M_d}{\Delta M_s}\right)_\text{exp}     & = 0.0285\pm0.0001\,, \\
 \left(\frac{\Delta M_d}{\Delta M_s}\right)_\text{Average} & = 0.0298_{-0.0009}^{+0.0005} = 0.0297_{-0.0003}^{+0.0003}\,(\text{had.})_{-0.0008}^{+0.0005}\,(\text{CKM})\,,
 \label{eq:DeltaMdOverDeltaMs}
\end{align}
where the parametric CKM uncertainty \eqref{eq:VtdVts} dominates theory error.
We note that there is a slightly discrepancy between our average and experiment, at the level of \(\sim \SI{1.4}{\sigma}\).
This allows a very 
precise determination of $|V_{td}/V_{ts}|$ from $B$ mixing~\cite{King:2019lal}. We note that the 
product of \eqref{eq:DeltaMdOverDeltaMs} and \eqref{eq:DeltaMsAVG2019} is about 3\% larger than 
the value for $\DeltaMd$ from a direct determination \eqref{eq:DeltaMdAVG2019}. The reason is that 
we also include the RBC/UKQCD~'18~\cite{Boyle:2018knm} result in the average \eqref{eq:xi_avg} for 
$\xi$ but not in the averages for the individual matrix elements since those were not determined 
there.\footnote{Furthermore the mean of a ratio is not the same as the ratio of the means.} 
The ellipses in Fig.~\ref{fig:DeltaMsVsDeltaMd} use $f_{B_s}^2B_1$ and $\xi$ as inputs 
and the ellipse for the weighted average therefore also gives a value of $\DeltaMd$ which is about 
3\% larger than \eqref{eq:DeltaMdAVG2019}. We further note that our ellipses for the FLAG~'19 and 
HPQCD~'19 results differ from the numerical results given in \cite{Bazavov:2016nty} and 
\cite{Dowdall:2019bea} because we use different values for the CKM parameters as discussed in 
\ref{sec:inputs}.

All in all we observe a consistent picture of different non-perturbative determinations for $B$ mixing.
The future ($\sim$~2025) prospects for lattice simulations of the matrix elements have been assessed in 
\cite{Kou:2018nap,Cerri:2018ypt}. The relative uncertainties for $f_{B_s}^2B_1$ are estimated as 
3\%~\cite{Kou:2018nap} and 1.3\%~\cite{Cerri:2018ypt}. Given that the two latest lattice results 
\cite{Bazavov:2016nty,Dowdall:2019bea} do not overlap within their uncertainties and differ by 15\% 
we believe the latter estimate to be overly optimistic and assume the scenario of~\cite{Kou:2018nap}. 
On the other hand the uncertainties in sum rule analyses are dominated by the one from the matching 
between QCD and Heavy Quark Effective Theory (HQET) matrix elements which can be reduced significantly by a perturbative two-loop 
matching calculation~\cite{Grozin:2017uto,Grozin:2018wtg} or eliminated altogether by using QCD~\cite{Korner:2003zk,Mannel:2007am}
instead of HQET sum rules. We expect that a reduction of the sum rule uncertainty for $B_1$ to 3\% 
is realistic. The effect on the weighted average is a reduction of the error from 3.1\% to about 2\%.

Simultaneously, new experimental data from Belle II and the LHC will improve our knowledge of the CKM 
element \V{cb} which constrains \V{ts} in the CKM unitarity fits and is the other critical input 
to our theory predictions. The Belle II collaboration forecasts \cite{Kou:2018nap} that by 2025 the 
uncertainty on \V{cb} will be at the level of 1\%. Using these future predictions (our 
explicit future inputs are given in \cref{tab:future_params}, with all others staying the same as at 
present), we expect that \(\DeltaMs\) will be predicted with an uncertainty of only \SI{\pm0.5}{\invps} 
by 2025, yielding \cref{eq:DeltaMsFuture}.\footnote{
It is worth noting that with such a precision, the FLAG 2019 central value given in \cref{eq:DeltaMsFLAG2019} 
is almost \SI{5}{\sigma} away from the experimental result.}

\begin{table}
\begin{tabular}{@{}LLL@{}}
\toprule
\text{Parameter} & \text{Future value} \\
\midrule
\V{cb} & 0.04240\pm 1\% = 0.04240 \pm 0.00042 \\
f_{\PBs} \sqrt{\hat{B}_1} & \SI[parse-numbers=false]{(262 \pm 1\%)}{\MeV} = \SI{262 \pm 2.6}{\MeV} \\
\bottomrule
\end{tabular}
\caption{
Future values for the two most important parameters for the error on $\DeltaMs$.
}
\label{tab:future_params}
\end{table}

\section{\texorpdfstring{\(\DeltaMs\)}{DeltaMs} interplay with flavour anomalies\label{BSM}}
\subsection{Status and future prospects of the \texorpdfstring{$b\to s\ell\ell$}{b to sll} anomalies\label{sec:anomalies}}

A key application of the refined SM prediction for $\DeltaMs$ is in the context of the recent 
hints of lepton flavour universality (LFU) violation in semi-leptonic \PB meson decays. 
Focussing on neutral current anomalies, the main observables are the LFU violating ratios 
$R_{K^{(*)}} \equiv \mathcal{B}(B \to K^{(*)} \mu^+ \mu^-) / \mathcal{B}(B \to K^{(*)} e^+ e^-)$ 
\cite{Aaij:2014ora,Aaij:2017vbb,Aaij:2019wad}, together with the angular distributions of 
$B \to K^{(*)} \mu^+ \mu^-$ \cite{Aaij:2015esa,Khachatryan:2015isa,Lees:2015ymt,Wei:2009zv,Aaltonen:2011ja,Aaij:2015oid,Abdesselam:2016llu,Wehle:2016yoi,Sirunyan:2017dhj,Aaboud:2018krd} 
and the branching ratios of hadronic $b \to s \mu^+ \mu^-$ decays \cite{Aaij:2014pli,Aaij:2015esa,Khachatryan:2015isa}. 
The effective Lagrangian for semi-leptonic $b \to s \mu^+ \mu^-$ transitions contains the terms
\begin{equation}
\label{eq:Leffbsmumu}
\Lagrangian^\text{NP}_{b \to s \mu \mu} \supset \frac{4 G_F}{\sqrt{2}} \V{tb} \V*{ts} \left( \delta C^\mu_9 O^\mu_9 + \delta C^\mu_{10} O^\mu_{10} + \delta C^{\prime\mu}_9 O^{\prime\mu}_9 + \delta C^{\prime\mu}_{10} O^{\prime\mu}_{10} \right) + \text{h.c.} \,,
\end{equation}
with 
\begin{align}
O^\mu_9 &= \frac{\alpha}{4 \pi} (\APqs_L \gamma_\mu \Pqb_L) (\bar \mu \gamma^\mu \mu) \, , \\
O^\mu_{10} &= \frac{\alpha}{4 \pi} (\APqs_L \gamma_\mu \Pqb_L) (\bar \mu \gamma^\mu \gamma_5 \mu) \,,
\end{align}
and the primed operators $O^{\prime \mu}_{9,10}$ having the opposite ($L \to R$) quark chiralities.
Intriguingly, one obtains a very good description of the $b\to s\ell\ell$ data by allowing 
new physics only in a single combination of Wilson coefficients (for recent fits see e.g.~\cite{Aebischer:2019mlg,Alguero:2019ptt,Kowalska:2019ley,Ciuchini:2019usw,Arnan:2019uhr,DAmico:2017mtc,Alok:2019ufo}). 
The best-fit scenario $\delta C_9^\mu = -\delta C_{10}^\mu \approx -0.53\pm0.08$ of 
\cite{Aebischer:2019mlg} corresponds to an effective operator with left-handed (LH) quark 
and muon currents and we consider its interplay with $\DeltaMs$ in Section~\ref{sec:interplay}. 
The data is also well represented by assuming either $\delta C_9^\mu$ or $\delta C_{10}^\mu$ 
being different from zero, corresponding to vector or axial-vector muon currents, respectively. 
However, we find that the mixing constraints on these scenarios are stronger than for case 
of LH muon currents, since larger values of the Wilson coefficients are required, 
and do not discuss them further. Following the update of $\RK$ from LHCb \cite{Aaij:2019wad}, 
there has been an increased favourability of right handed (RH) current contributions in the 
quark sector and we investigate their effect in Section~\ref{sub:Zprime_RH}. 

By 2025 LHCb expects a serious improvement in the measured precision on \(\RK\) and \(\RKstar\) 
compared to the situation today -- they forecast \cite{Bediaga:2018lhg} the error on \(\RK\) 
will be reduced to \(\pm 0.025\) (better than a factor of two compared to the error reported at 
Moriond 2019) and to \(\pm 0.031\) for \(\RKstar\) (a factor of four better).
If these precisions are indeed realised, and the central values stay close to their currently 
measured values, LHCb will have made a discovery of LFU violation at the level of 
\SI{6}{\sigma} and \SI{10}{\sigma} for \(\RK\) and \(\RKRKstar\) respectively.

\subsection{New physics contributions to \texorpdfstring{$\DeltaMs$}{DeltaMs} in the effective theory\label{sec:MixingBSM}}

The NP contributions to $B_s$ mixing can be described by the effective Lagrangian
\begin{align}
\label{eq:LeffBmixing}
\Lagrangian^\text{NP}_{\Delta B = 2} &\supset - \frac{4 G_F}{\sqrt{2}} \left( \V{tb} \V*{ts} \right)^2 \Big[ 
C^{LL}_{bs} \left( \APqs_L \gamma_\mu \Pqb_L \right)^2 
+ C^{RR}_{bs} \left( \APqs_R \gamma_\mu \Pqb_R \right)^2 \nonumber \\ 
& + C^{LR}_{bs} \left( \APqs_L \gamma_\mu \Pqb_L \right) \left( \APqs_R \gamma^\mu \Pqb_R \right) \Big] + \text{h.c.} \,,
\end{align}
where we assume NP only gives rise to vector colour singlet operators.
The full basis of \(\Delta B = 2\) operators (\cref{eq:operator_basis_Q1,eq:operator_basis_Q25}) also contains operators 
that give tensor or scalar Dirac structures when written in colour singlet form. However, these 
operator structures are highly disfavoured by the fits to \(b \to s \ell \ell\) data, and so we ignore them here.
Assuming the NP coefficients $C_{bs}$ are generated at some higher scale $\mu_\text{NP}$, we have 
to include RG running effects down to the \Pqb quark scale (see e.g.~\cite{Bagger:1997gg}), which 
brings in the LR vector operator with the non colour singlet structure through operator mixing 
and explains the appearance of both \(B_4\) and \(B_5\) in the expression below. In this way we 
can parameterise the SM+NP contribution normalized to the SM one as\footnote{This expression neglects 
the top-quark threshold, which is a sub-percent effect. However, the correct running is taken into 
account in our numerics and figures.}
\begin{align}
\label{eq:BmixingLLRRLR}
\frac{\DeltaMs^\text{SM+NP}}{\DeltaMs^\text{SM}} = 
\Bigg| 1 &+ \frac{\eta^{6/23}}{R_\text{loop}^\text{SM}} 
\Bigg\{ C^{LL}_{bs} + C^{RR}_{bs} 
- \frac{C^{LR}_{bs}}{2\eta^{3/23}} \Bigg[\frac{B_5}{B_1} 
\left( \frac{M_{\PBs}^2}{(m_b + m_s)^2} + \frac{3}{2} \right) \nonumber \\
&
+ \frac{B_4}{B_1} \left( \frac{M_{\PBs}^2}{(m_b + m_s)^2} + \frac{1}{6} \right) 
\left(\eta^{-27/23} - 1\right) \Bigg] \Bigg\} \Bigg| \, ,
\end{align}
with $\eta = \alpha_s(\mu_\text{NP}) / \alpha_s(m_b)$, the bag parameters $B_i$ defined as in 
\cref{defB1,eq:operator_basis_bags}, and the SM loop function given by 
\begin{equation}
R^\text{loop}_\text{SM} = \frac{\sqrt{2} G_F M_W^2 \hat\eta_B S_0(x_t)}{16 \pi^2} = (1.310 \pm 0.010) \times 10^{-3} \,.
\end{equation}
In general, one would expect that any NP contributing to $b \to s \ell\ell$ transitions will eventually 
feed into $B_s$ mixing. However, this connection is hidden at the effective operator level where the 
Wilson coefficients in Eq.~\eqref{eq:Leffbsmumu} and Eq.~\eqref{eq:BmixingLLRRLR} are independent.\footnote{The double insertion of the effective Lagrangian for $b\to s\ell\ell$ 
transitions \eqref{eq:Leffbsmumu} yields a contribution to $\DeltaMs$ at the dimension-eight level. Compared to the contribution 
from Eq.~\eqref{eq:BmixingLLRRLR} this is suppressed by $m_b^2/\Lambda_\text{NP}^2$ and cannot be used to 
obtain meaningful constraints.} 
It is hence crucial to focus on specific UV realizations in order to explore the connection between 
$b \to s \ell\ell$ anomalies and $\Delta M_s$ and we will consider two simplified models below.

\subsection{Simplified models for the \texorpdfstring{$b\to s\ell\ell$}{b to sll} anomalies\label{sub:Zprime_intro}}

There are two basic possibilities how the effective operators in Eq.~\eqref{eq:Leffbsmumu} 
can be generated at tree level: by the exchange of a $Z'$ mediator coupling to the quark 
and the muon current or by the exchange of a lepto-quark coupling to mixed quark-muon 
currents. Anticipating our results from Section~\ref{sec:interplay} we describe the $Z'$ 
scenario in detail below. In addition we consider the case of the scalar lepto-quark 
$S_3 \sim (3,3,1/3)$, whose defining Lagrangian can be found e.g.~in \cite{DiLuzio:2017fdq,Dorsner:2016wpm}. 

We consider a simplified \PZprime model with only those couplings required to explain the 
observed flavour anomalies, with the Lagrangian 
\begin{equation}
\label{eq:LZp}
\Lagrangian_{\PZprime} \supset \frac{1}{2} M_{\PZprime}^2 (Z'_\mu)^2 
+ \PZprime_\mu \left( \lambda^Q_{ij} \APqd_L^i \gamma^\mu \Pqd_L^i 
+ \lambda^d_{ij} \APqd_R^i \gamma^\mu \Pqd_R^j 
+ \lambda^L_{ij} \APlepton_L^i \gamma^\mu \Plepton_L^j \right) \,,
\end{equation}
where $\Pqd^i$ and $\Plepton^i$ label the different generations of down-type quark and charged 
lepton mass eigenstates respectively, and $\lambda^{Q,d,L}$ are hermitian flavour space matrices. 
We have neglected RH currents in the lepton sector in the above Eq.~(\ref{eq:LZp}) 
since the latter actually worsen the compatibility with $\Delta M_s$, as they require a larger 
Wilson coefficient. 

Integrating out the \PZprime at tree level, we get the effective Lagrangian 
\begin{align}
\Lagrangian^\text{eff}_{\PZprime} \supset -\frac{1}{2 M_{\PZprime}^2} \bigg(&2 \lambda^Q_{23} \lambda^L_{22} (\APqs_L \gamma^\mu \Pqb_L)(\bar{\mu}_L \gamma_\mu \mu_L) + 2 \lambda^d_{23} \lambda^L_{22} (\APqs_R \gamma^\mu \Pqb_R)(\bar{\mu}_L \gamma_\mu \mu_L) \nonumber \\
&+ \left(\lambda^Q_{23}\right)^2 (\APqs_L \gamma^\mu \Pqb_L)(\APqs_L \gamma_\mu \Pqb_L) \\
&+ \left(\lambda^d_{23}\right)^2 (\APqs_R \gamma^\mu \Pqb_R)(\APqs_R \gamma_\mu \Pqb_R) + 2 \lambda^Q_{23} \lambda^d_{23} (\APqs_L \gamma^\mu \Pqb_L)(\APqs_R \gamma_\mu \Pqb_R) \bigg) \,. \nonumber
\end{align}
The first line contains the terms that contribute to the rare decay coefficients $C_{9,10}^{(\prime) \mu}$, 
the second a contribution to $\DeltaMs$ through the same operator as in the SM, 
and the third contributions to $\DeltaMs$ from operators that do not appear in the SM.

Matching onto the effective Lagrangians for the low energy observables in 
\cref{eq:Leffbsmumu,eq:LeffBmixing}, we find (at the scale $\mu = M_{\PZprime}$)
\begin{align}
\delta C_9^\mu = -\delta C_{10}^\mu &= - \frac{\pi}{\sqrt{2} G_F M^2_{\PZprime} \alpha} \left( \frac{\lambda^Q_{23} \lambda^L_{22}}{\V{tb} \V*{ts}} \right) \,, \\
\delta C_9^{\prime \mu} = -\delta C_{10}^{\prime \mu} &= - \frac{\pi}{\sqrt{2} G_F M^2_{\PZprime} \alpha} \left( \frac{\lambda^d_{23} \lambda^L_{22}}{\V{tb} \V*{ts}} \right) \,, \\
C_{bs}^{LL} &= \frac{1}{4 \sqrt{2} G_F M^2_{\PZprime}} \left( \frac{\lambda^Q_{23}}{\V{tb} \V*{ts}} \right)^2 \,, \\
C_{bs}^{RR} &= \frac{1}{4 \sqrt{2} G_F M^2_{\PZprime}} \left( \frac{\lambda^d_{23}}{\V{tb} \V*{ts}} \right)^2 \,, \\
C_{bs}^{LR} &= \frac{\sqrt{2}}{4 G_F M^2_{\PZprime}} \left( \frac{\lambda^Q_{23} \lambda^d_{23}}{(\V{tb} \V*{ts})^2} \right) \,.
\end{align}
Inserting the above expressions in \cref{eq:BmixingLLRRLR}, using our combined averages for 
the bag parameters, and taking a typical \PZprime scale of \SI{5}{\TeV} we find that the contribution to $\DeltaMs$ can be written as\footnote{The prefactor and 
the relative size of the interference term have a sub-leading logarithmic dependence from the 
\PZprime mass (less than 1\% for $M_{Z'}$ in the \SIrange{1}{10}{\TeV} range).}
\begin{equation}
\label{eq:Zprime5TeVmixingestimate}
\frac{\DeltaMs^\text{SM+NP}}{\DeltaMs^\text{SM}} \approx \left| 1 + 200 \left(\frac{\SI{5}{\TeV}}{M_{Z'}}\right)^2 
\left[ \left(\lambda^Q_{23}\right)^2 + \left(\lambda^d_{23}\right)^2 - 9 \lambda^Q_{23} \lambda^d_{23} \right] \right| \,, 
\end{equation}
which shows a significant RG enhancement for the LR operator.  

Besides $B_s$ mixing, we consider a constraint that is particularly relevant for light $Z'$ 
interacting with muons via LH currents. By $SU(2)_L$ invariance the $Z'$ couples also to LH 
neutrinos via the $\lambda^L_{22}$ coupling which is required by the $b \to s \mu\mu$ anomalies. 
Hence, one has an extra term in the $Z'$ effective Lagrangian of the type
\begin{equation}
\Lagrangian^\text{eff}_{\PZprime} 
\supset -\frac{(\lambda^L_{22})^2}{M_{Z'}^2} (\bar \mu_L \gamma^\mu \mu_L) 
(\bar \nu_{\mu L} \gamma^\mu \nu_{\mu L}) \, ,
\end{equation}
which leads to the neutrino trident production, $\nu_\mu N \to \nu_\mu \mu^+ \mu^- N$. 
Using the most recent calculation for the SM cross-section, one gets \cite{Altmannshofer:2019zhy} 
\begin{equation}
\label{eq:tridCCFR}
\frac{\sigma^{\rm SM+NP}_{\rm CCFR}}{\sigma^{\rm SM}_{\rm CCFR}} 
= \frac{1.13 \left(1 + \frac{v^2 (\lambda^L_{22})^2}{M^2_{Z'}}\right)^2
+ \left( 1+ 4 s^2_W + \frac{v^2 (\lambda^L_{22})^2}{M^2_{Z'}} \right)^2}{1.13+(1+4 s^2_W)^2} 
\end{equation}
(with $v\approx246$ GeV and $s^2_W \approx 0.231$), which is constrained by the existing 
CCFR measurement $\sigma_{\rm CCFR} / \sigma^{\rm SM}_{\rm CCFR} = 0.82 \pm 0.28$ \cite{Mishra:1991bv}.
This result implies \(M_{Z'} / \lambda^L_{22} > \SI{0.47}{\TeV}\) at \SI{2}{\sigma}. The upcoming 
DUNE experiment \cite{Abi:2018dnh} is expected to also measure this process, however 
the precision it will achieve (see e.g.~\cite{Altmannshofer:2019zhy}), suggest its data will not 
increase much the limits on this parameter combination and so we will use the CCFR bound throughout.

Last but not least, the parameter space is constrained by direct LHC searches~\cite{Aaboud:2017buh,Sirunyan:2017yrk} 
and perturbative unitarity~\cite{DiLuzio:2017chi,DiLuzio:2016sur}. Whether a few TeV $Z'$ is ruled 
out by LHC direct searches crucially depends on the details of the $Z'$ model. The stringent 
constraints from di-lepton searches \cite{Aaboud:2017buh} are tamed in models where the $Z'$ does 
not couple to valence quarks. For instance, assuming only the two couplings required by the 
$b \to s \ell\ell$ anomaly, namely $\lambda^Q_{23}$ and $\lambda^L_{22}$,  one finds that current 
$Z'$ searches are not sensitive yet (see e.g.~\cite{Allanach:2017bta,Afik:2018nlr}).
Assuming the best-fit scenario $\delta C_9^\mu = -\delta C_{10}^\mu \approx -0.53\pm0.08$ 
\cite{Aebischer:2019mlg} is generated by the exchange of a $Z'$ we obtain the perturbative unitarity 
bound $M_{Z'} < \SI{59}{\TeV}$ \cite{DiLuzio:2017chi}. The same limit for the lepto-quark scenario reads $M_{S_3} < \SI{69}{\TeV}$ \cite{DiLuzio:2017chi}. 
The mass of the lepto-quark is also bounded from below up to about \SI{1}{\TeV} \cite{Sirunyan:2017yrk} 
from direct searches at the LHC.

\subsection{\texorpdfstring{$\DeltaMs$}{DeltaMs} interplay with flavour anomalies in the minimal scenarios\label{sec:interplay}}

We investigate which new physics contributions $\DeltaMs^\text{NP}$ to the $B_s$ mass difference 
are generated by the simplified models of Section~\ref{sub:Zprime_intro}, assuming 
LH currents in the quark sector and real couplings (these two assumptions will be relaxed in 
Section~\ref{sec:ZprimeOptions}). We obtain the predictions 
\begin{align}
\label{eq:DeltaMs_C9_dependency_Zprime}
\frac{\DeltaMs^\text{SM+NP}}{\DeltaMs^\text{SM}} &= \left| 1 + \frac{G_F \alpha_\text{EM}^2 \eta^{6/23}}{2\sqrt{2} \pi^2 R_\text{loop}^\text{SM}} 
   \left(\delta C_9^\mu = - \delta C_{10}^\mu\right)^2 \left(\frac{M_{\PZprime}}{\lambda^L_{22}}\right)^2 \right| \\
&\approx \left| 1 + \frac{1}{360} \left(\frac{\delta C_9^\mu = - \delta C_{10}^\mu}{-0.53}\right)^2 \left(\frac{M_{\PZprime}/\lambda^L_{22}}{\SI{1}{\TeV}}\right)^2 \right| \, , \nonumber
\end{align}
for the $Z'$ and 
\begin{align}
\label{eq:DeltaMs_C9_dependency_LQ}
\frac{\DeltaMs^\text{SM+NP}}{\DeltaMs^\text{SM}} &= \left| 1 + \frac{G_F \alpha_\text{EM}^2 \eta^{6/23}}{2\sqrt{2} \pi^2 R_\text{loop}^\text{SM}} \frac{5}{64 \pi^2} 
    \left(\delta C_9^\mu = - \delta C_{10}^\mu\right)^2 \left(M_{S_3}\right)^2 \right| \\
&\approx \left| 1 + \frac{1}{\num{29000}} \left(\frac{\delta C_9^\mu = - \delta C_{10}^\mu}{-0.53}\right)^2 \left(\frac{M_{S_3}}{\SI{1}{\TeV}}\right)^2 \right|\, , \nonumber
\end{align}
for the lepto-quark. 

\begin{figure}[th]
\includegraphics[height=0.47\textwidth]{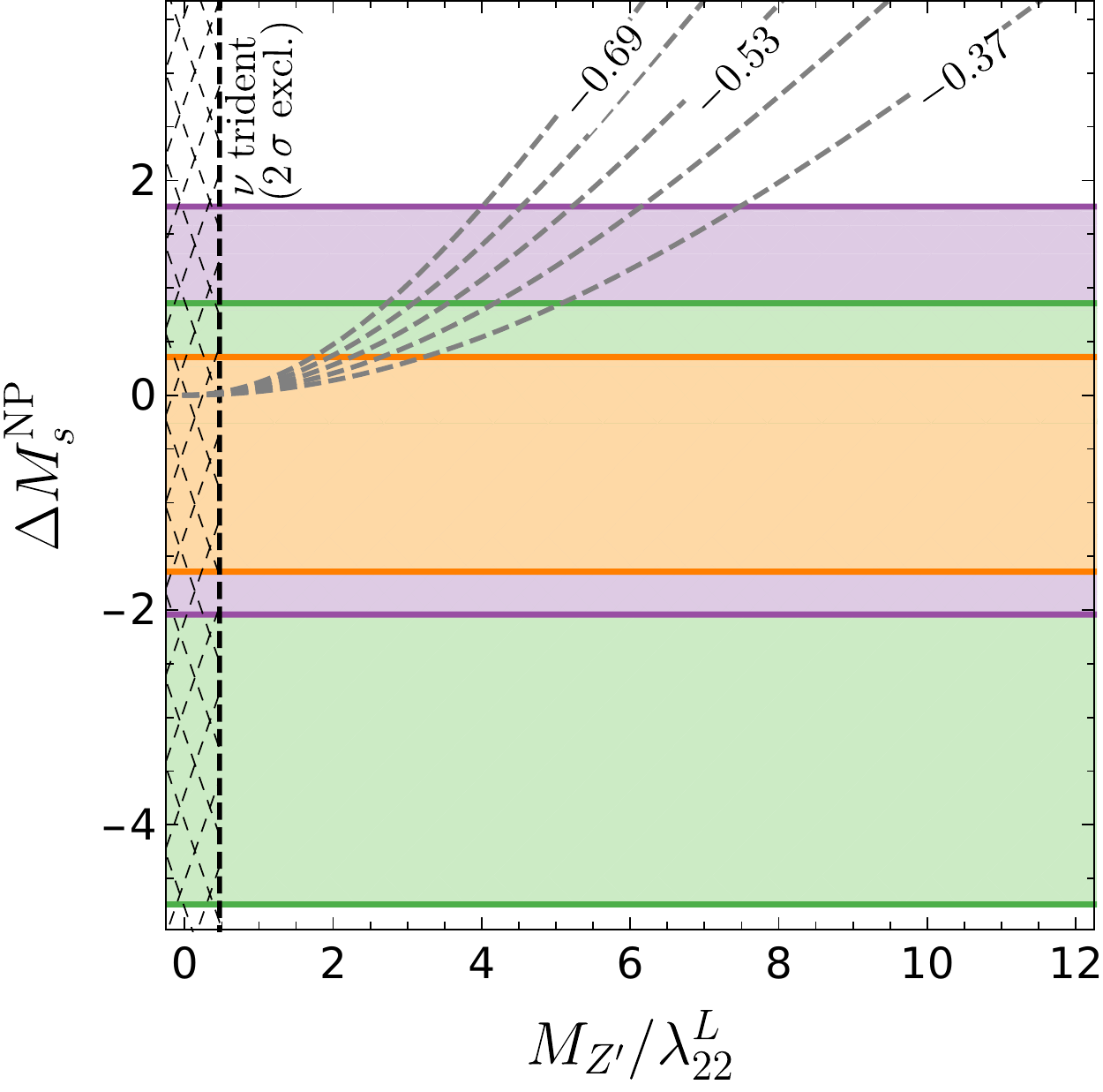}
\hfill
\includegraphics[height=0.47\textwidth]{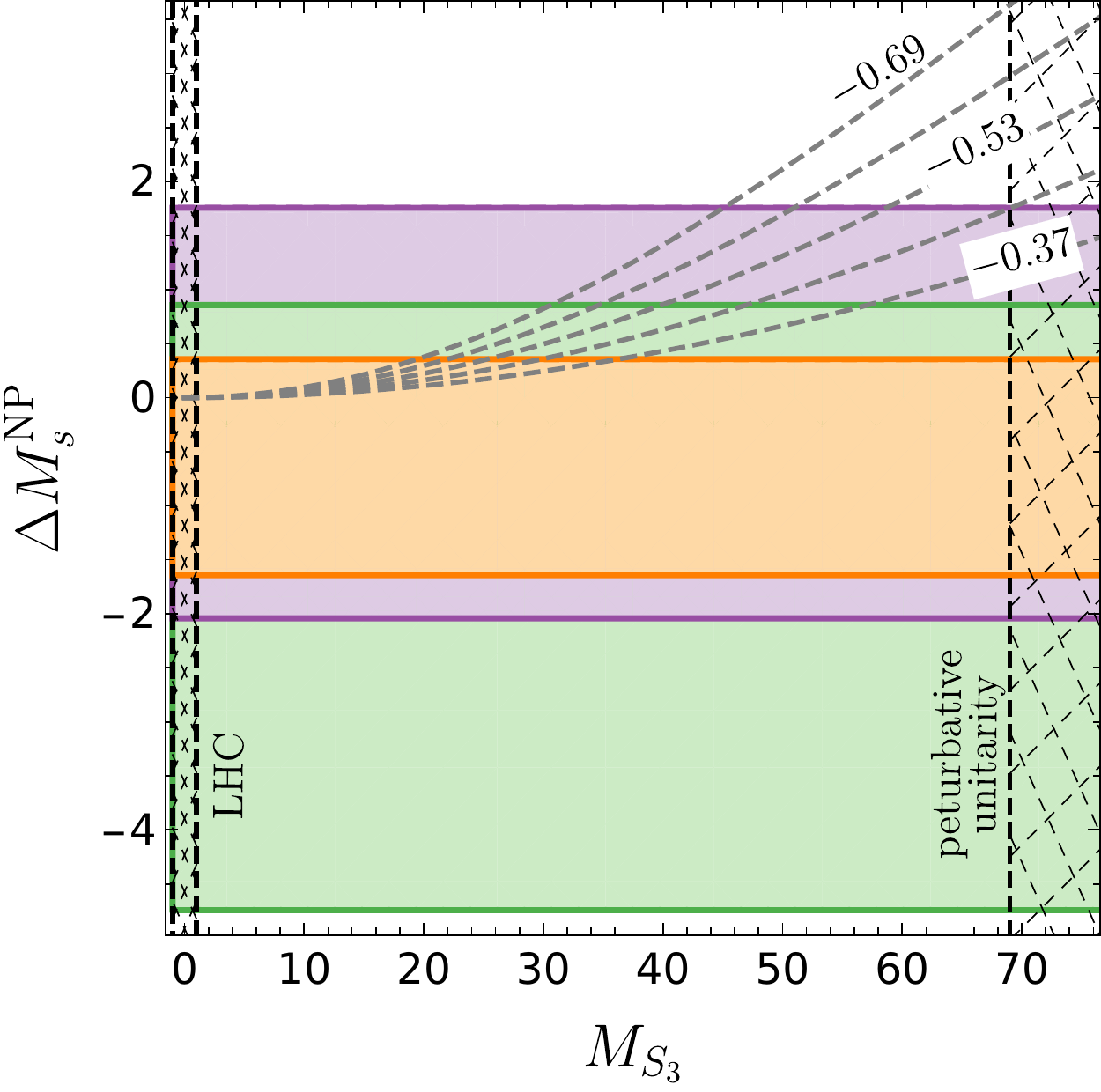}
\includegraphics{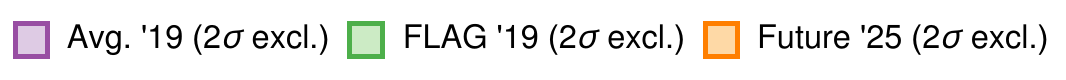}
\caption{We show the predicted new physics effects $\Delta M_s^{\rm NP}$ 
(dashed lines) for various values of $\delta C_9^\mu = - \delta C_{10}^\mu$ 
in comparison with the allowed two-sigma regions (shaded) in the FLAG~'19, 
Avg.~'19 and Future~'25 scenarios. The left and right panel assume the 
simplified $Z'$ model and scalar lepto-quark model, respectively. 
Hatched regions show limits from neutrino trident production or direct 
LHC searches and perturbative unitarity.}
\label{fig:deltaMSvsZLQfuture}
\end{figure}

In \cref{fig:deltaMSvsZLQfuture} we compare the new physics contributions 
$\Delta M_s^\text{NP}$ in the two models to the allowed two-sigma regions for the 
present and future scenarios. The dashed lines show what effect is predicted for a 
given value of the Wilson coefficient combination $\delta C_9^\mu = - \delta C_{10}^\mu$, 
whereby the envelope of these lines corresponds to the two-sigma region from the fit 
$\delta C_9^\mu = - \delta C_{10}^\mu = -0.53\pm0.08$ to the present $b\to s\ell\ell$ data. 
In addition we show constraints from neutrino trident production, direct searches by the 
LHC experiments and perturbative unitarity as hatched regions. 

The left plot shows the simplified $Z'$ model where the relation between $\Delta M_s$ and 
$\delta C_9$ depends on the parameter combination $M_{\PZprime}/\lambda^L_{22}$. 
We observe that a $Z'$ explanation of the $b\to s\ell\ell$ data causes a sizeable positive 
contribution to $\DeltaMs$ unless the $Z'$ mass is rather light. Values of $\DeltaMs^\text{NP}$ 
outside the shaded regions are in tension with the respective predictions for $\DeltaMs$ at 
the level of two standard deviations. This allows us to place upper bounds on the mass of 
the $Z'$.\footnote{A full statistical fit would be in order here to make a definitive 
statement about some particular model being excluded.}  For instance, we find that the large 
central value of the FLAG~'19 prediction in Eq.~\eqref{eq:DeltaMsFLAG2019} (green region) 
requires $M_Z' \lesssim \SI{4}{\TeV}$ for $\lambda^L_{22}=1$ in order to explain $b \to s \ell \ell $ 
at \SI{1}{\sigma}. 
The limit is about a factor $\sfrac{3}{2}$ weaker in the case of the Average~'19 value of 
$\Delta M_s$ (purple region) despite the reduction of the uncertainty by over 30\% with 
respect to FLAG~'19 because the central value is closer to the experimental result. 
It is straightforward to rescale the bound for different lepton couplings. The combination 
of $\DeltaMs$ and perturbative unitarity (requiring $\lambda^L_{22}<\sqrt{4\pi}$ 
\cite{DiLuzio:2017chi,DiLuzio:2016sur}) gives an upper bound of $M_{Z'}\lesssim \SI{20}{\TeV}$ 
from the Average~'19 prediction. With the future scenario for $\Delta M_s$ (orange region) 
this will improve to about \SI{9}{\TeV}, being a factor of 6 more constraining than the perturbative 
unitarity bound \cite{DiLuzio:2017chi,DiLuzio:2016sur}. 

The lepto-quark model is shown in the right plot and yields smaller new physics contributions 
to $\DeltaMs$ because the effect is only generated at loop level. Using the Average~'19 value 
(purple) we find that a lepto-quark explanation is still viable in the entire mass range between 
the LHC and unitarity constraints. However, the increased precision of the Future '25 scenario 
implies an upper limit of about \SI{30}{\TeV} on the lepto-quark mass $M_{S_3}$, better than the 
perturbativity bound by more than a factor of two. 

We conclude that both simplified $Z'$ and lepto-quark models currently provide possible solutions 
to the $b\to s\ell\ell$ anomalies, with the previous stringent constraints \cite{DiLuzio:2017fdq} 
from the FLAG~'19 values for $\DeltaMs$ being relaxed by the refined SM prediction Average~'19.  
However, with the future increase in the precision of $\DeltaMs$ lepto-quark explanations will be 
favoured over a $Z'$, assuming that the central values remain similar. In Section~\ref{sec:ZprimeOptions} 
we therefore discuss ideas how the implied tension might be reduced by generalizations of the 
minimal $Z'$ scenario.

\subsection{New physics options beyond the minimal $Z'$ scenario\label{sec:ZprimeOptions}}

The constraints on the minimal $Z'$ scenario considered above 
are particularly strong because the latter predicts a 
\emph{positive} contribution to $\DeltaMs$, while the central values of the current SM predictions 
are already larger than the experimental result. Therefore, we will explore two simple options 
for how a negative contribution to $\DeltaMs$ could be generated, thus relaxing the tension. 
Examining \cref{eq:Zprime5TeVmixingestimate} the two cases present themselves:
\begin{enumerate}
	\item For only LH quark coupling (i.e.~$\lambda^d = 0$), a negative contribution can only be generated if $\lambda^Q_{23}$ acquires a complex phase.
	\item With both LH and RH quark couplings, real couplings alone can give the required sign if the interference term is large enough.
\end{enumerate}
Although these two possibilities were already identified in \cite{DiLuzio:2017fdq}, 
their compatibility with global fits for flavour anomalies was only partially 
assessed in \cite{DiLuzio:2018wch}. We start with CP violating couplings in 
\cref{sub:Zprime_complex}, and then study the case for RH quark currents in \cref{sub:Zprime_RH}. 
As a benchmark point, we take a lepton coupling of $\lambda^L_{22} = 1$ and a \PZprime mass 
of \SI{5}{\TeV} where the best-fit result $\delta C_9^\mu = - \delta C_{10}^\mu=-0.53$ saturates 
the two-sigma range of the Average~'19 prediction (see \cref{fig:deltaMSvsZLQfuture}). 
In the following, we address the question whether either option 1.\ or 2.\ allows for this 
benchmark point to be viable within the FLAG~'19 or Future~'25 scenarios.

\subsubsection{CP violating couplings\label{sub:Zprime_complex}}

It is clear from \cref{eq:Zprime5TeVmixingestimate} that if $\lambda^Q_{23}$ has a large enough imaginary part, then it will be possible to bring down the theory prediction for $\DeltaMs$
below the SM prediction and into better agreement with experiment.
However, once we allow for complex \PZprime quark couplings, there are extra constraints to be considered, in the form of CP-violating observables that arise from \PBs mixing.
The most relevant here is the mixing-induced CP asymmetry \cite{Lenz:2006hd,Artuso:2015swg}, arising from interference between $B$ meson mixing and decay. 
The semi-leptonic CP asymmetries for flavour-specific decays, $a^s_\text{sl}$, are not 
competitive yet since the experimental errors are still too large \cite{Artuso:2015swg}.
Defining 
\begin{equation} 
\phi_{\Delta} = \text{Arg} \left( 1 + \frac{\eta^{6/23} C^{LL}_{bs}}{R^{\rm loop}_{\rm SM}} \right) \, ,
\end{equation}
(with \(\eta\) as in \cref{eq:BmixingLLRRLR}) the mixing-induced CP asymmetry is given by
\begin{equation}
A^\text{mix}_\text{CP} (\PBs \to \PJpsi \Pphi) = \sin{\left( \phi_\Delta - 2 \beta_s \right)} \, ,
\end{equation}
where $A^\text{mix}_\text{CP} = -0.021 \pm 0.031$ \cite{Amhis:2016xyh,HFLAV:PDG2018}, $\beta_s = 0.01843^{+0.00048}_{-0.00034}$ \cite{CKMfitter:Summer18}, and we have neglected penguin contributions \cite{Artuso:2015swg}.

New phases in $\lambda^Q_{23}$ also imply a complex value for $\delta C_9^\mu$, which has not typically been considered in previous global fits (see \cite{Alok:2017jgr,Alda:2018mfy} for exceptions).
We perform our own fit using the \texttt{flavio} software \cite{Straub:2018kue}, using the same set of \(b \to s \ell \ell\) observables as \cite{Altmannshofer:2017fio}.

\begin{figure}
\includegraphics[width=0.5\textwidth]{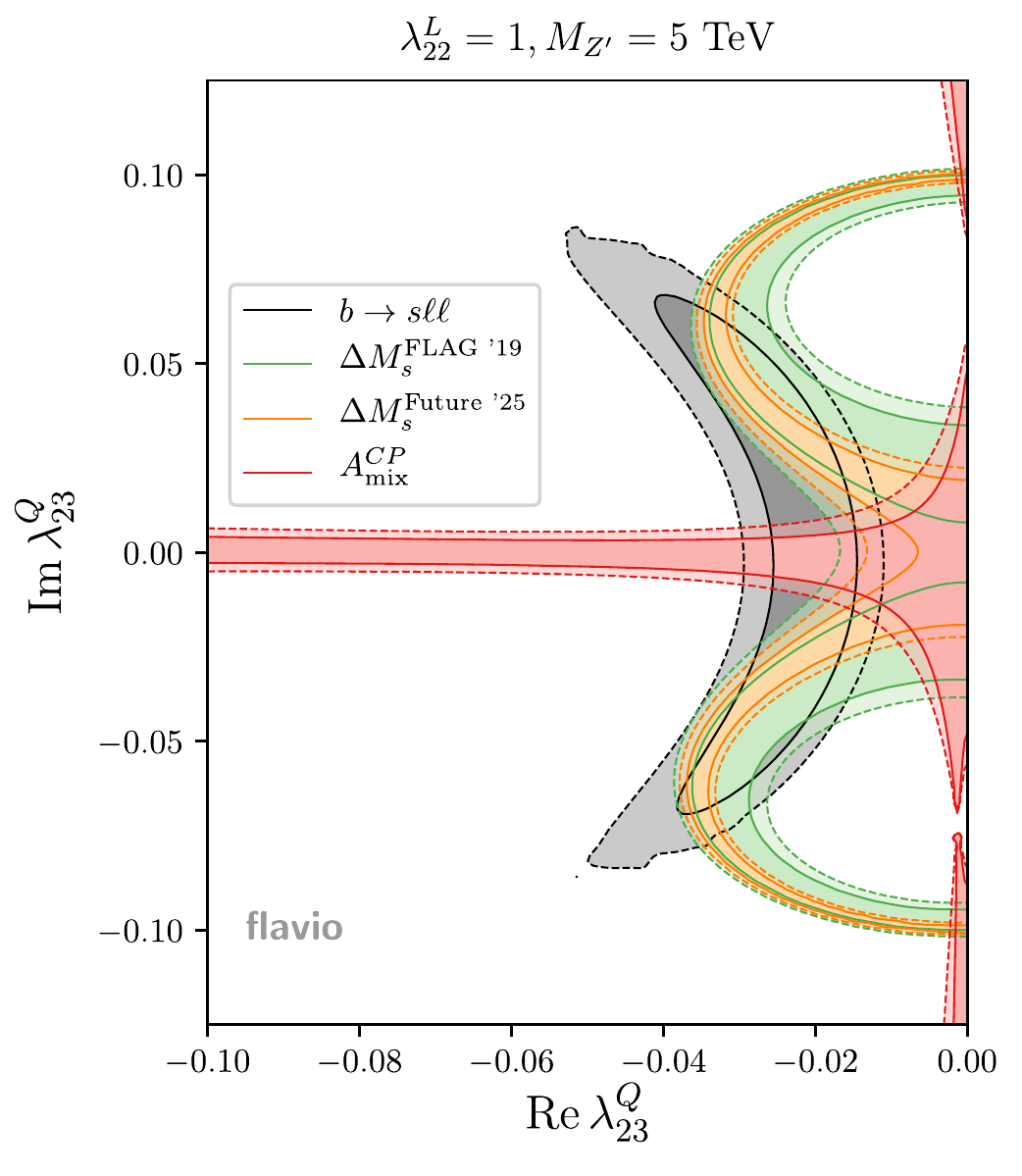}
\caption{Allowed regions at \SI{1}{\sigma} (solid) and \SI{2}{\sigma} (dashed) for $b \to s \ell \ell$ (grey), $\DeltaMs$ (green (FLAG~'19) and orange (Future~'25)), and $A^\text{CP}_\text{mix}$ (red).}
\label{fig:complex_Zprime}
\end{figure}

The result of our fit is shown in \cref{fig:complex_Zprime}, with the higher SM prediction $\DeltaMs^\text{FLAG '19}$ corresponding to the green bands.
There is no overlap between the \SI{1}{\sigma} regions for our three observables, since $A^\text{mix}_\text{CP}$ provides too strong a constraint on the imaginary part of the coupling.
So a complex coupling does not provide a way to significantly evade the strong bounds from $\DeltaMs$.
Using instead our prediction \(\DeltaMs^\text{Future '25}\) (orange) there is still no overlap, for the same reason.
As such, the addition of complex phases to the \PZprime quark coupling does not alleviate tension at the benchmark point in the Future~'25 scenario, which is important given the strength of this future bound, as shown in \cref{fig:deltaMSvsZLQfuture}.

\subsubsection{Right-handed couplings\label{sub:Zprime_RH}}

A \PZprime coupling to a RH quark current can also allow for a negative BSM contribution to $\DeltaMs$ through the interference term, which has the advantage that it is RG enhanced by roughly an order of magnitude (see \cref{eq:Zprime5TeVmixingestimate}) relative to the pure LL or RR operators.
In our previous work \cite{DiLuzio:2017fdq,DiLuzio:2018wch} we stated that RH quark currents break the approximate symmetry $\RK \approx \RKstar$ and hence were disfavoured by the data.
On the other hand, the recent update of $\RK$ at Moriond 2019 \cite{Aaij:2019wad} 
favours a non-zero RH quark current contribution, as widely discussed in the literature since then (see e.g.\ \cite{Aebischer:2019mlg,Alguero:2019ptt,Kowalska:2019ley,Ciuchini:2019usw,Arnan:2019uhr,DAmico:2017mtc,Alok:2019ufo}).

\begin{figure}
\includegraphics[width=0.6\textwidth]{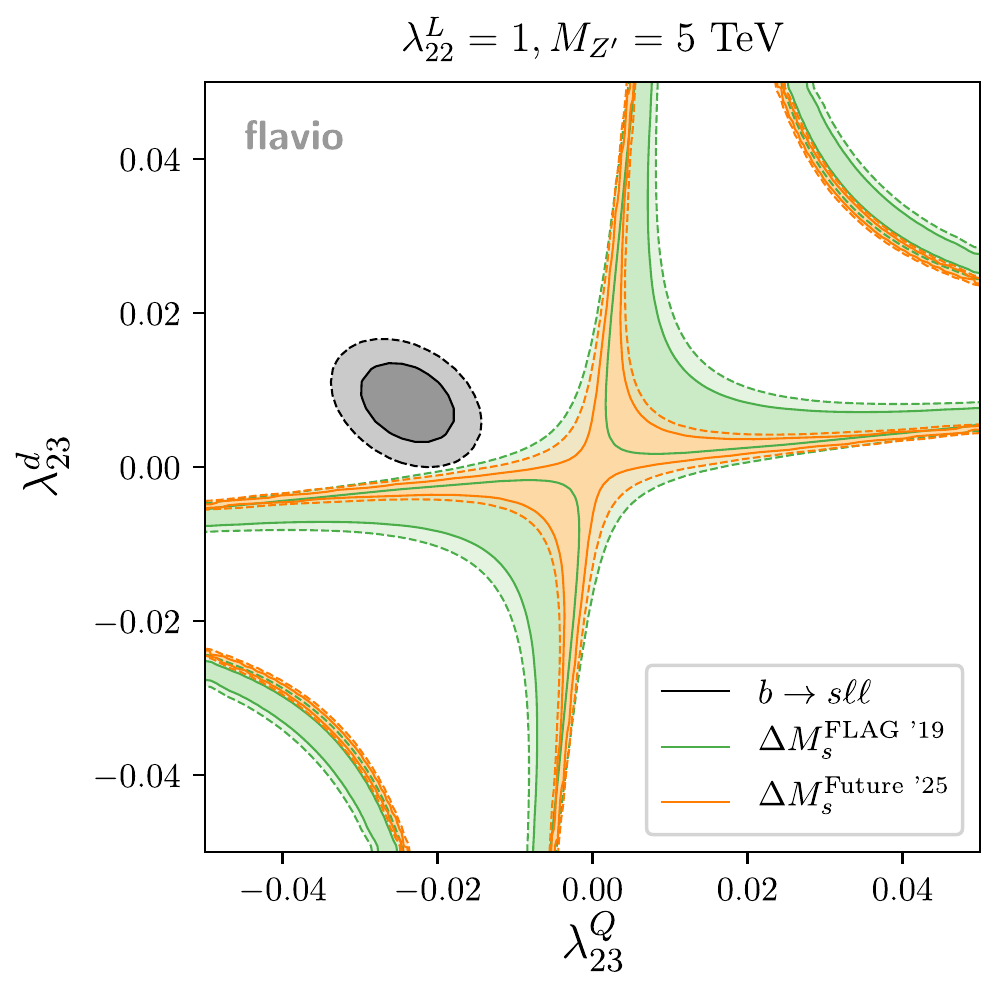}
\caption{
Allowed regions at \SI{1}{\sigma} (solid) and \SI{2}{\sigma} (dashed) for $b \to s \ell \ell$ (grey) and $\DeltaMs$ (green (FLAG~'19) and orange (Future~'25)).}
\label{fig:LR_Zprime}
\end{figure}

There is however a problem with this solution, 
which is that
in our \PZprime model, at the leading order in the NP contribution, 
the behaviour can be written as
\begin{equation}
\frac{R_K - 1}{R_{K^*}-1} \approx 
\frac{\lambda^Q_{23} + \lambda^d_{23}}{\lambda^Q_{23} - (2p-1) \lambda^d_{23}} \, ,
\end{equation}
where $p \approx 0.86$ is the polarization fraction \cite{Hiller:2014ula}. 
The current experimental measurements suggest 
$(R_K - 1) / (R_{K^*}-1) \approx 0.50$, 
which requires $\lambda^d_{23} / \lambda^Q_{23} \approx -0.37$.
On the other hand, \cref{eq:Zprime5TeVmixingestimate} shows that 
the same sign for LH and RH quark couplings are needed to reduce $\DeltaMs$.
This is evident in our fit result shown in \cref{fig:LR_Zprime}, where the allowed region for the \(b \to s \ell \ell\) anomalies is partially driven by the \(\RKRKstar\) measurements.
With the high prediction $\DeltaMs^\text{FLAG '19}$, there is a clear gap between the \SI{1}{\sigma} allowed regions, 
which does not allow us to solve the tension for the benchmark point 
(for a similar conclusion see also \cite{Arnan:2019uhr}).
If we instead examine the orange region, corresponding to the \(\DeltaMs^\text{Future '25}\) prediction,\footnote{For the numerics here we assume a similar relative improvement in the precision of the BSM bag parameters \(B_{4,5}\) as was described earlier for \(B_1\).} the gap has shrunk by only a small amount, and so our benchmark point is only marginally more favoured.

\subsubsection{Beyond simplified $Z'$ models\label{sub:beyondminZp}}

Within the simplified $Z'$ model considered so far, 
neither of the two options above 
(imaginary couplings and RH quark currents) 
helps to improve the compatibility between 
the $b \to s \ell \ell$ measurements and the expected precision determination 
value of $\Delta M_s$ in 2025. 
There are, however, other possibilities in order 
to achieve that when going beyond our simplified $Z'$ setup. 
For instance, sticking only to $R_{K}$ and $R_{K^*}$ 
(so no angular observables, etc.), 
these can be accommodated via NP 
in electrons featuring sizeable contributions 
both from LH and RH quark currents, 
so that a negative contribution to $\Delta M_s$ is possible 
as well. This class of models were suggested in the context 
of Composite Higgs scenarios \cite{Carmona:2017fsn}.  
Alternatively, 
as pointed out in \cite{Bordone:2018nbg}, in UV complete models 
of the vector lepto-quark $U_\mu \sim (3,1,2/3)$ \cite{DiLuzio:2017vat,Calibbi:2017qbu,Bordone:2017bld,Barbieri:2017tuq,Blanke:2018sro,DiLuzio:2018zxy} addressing both $R_{D^{(*)}}$ and $b \to s \ell \ell$ anomalies, 
the couplings to quarks of extra $Z'$ and/or coloron states 
(not directly responsible for the anomalies 
in semi-leptonic $B$ decays) can naturally have 
a large phase in order to accommodate a negative $\Delta M_s$, 
without being in tension with CP violating observables. 
These two examples show that although it is 
difficult to obtain $\Delta M_s^{\rm NP} < 0$ 
within simplified 
models for the $b \to s \ell \ell$ anomalies, 
that is certainly not impossible in more general constructions.

\section{Conclusions\label{conclusions}}

In this paper, we have combined recent results from sum rules~\cite{Kirk:2017juj,King:2019lal} 
(by some of the current authors, see also \cite{Grozin:2016uqy,Grozin:2017uto,Grozin:2018wtg} 
for consistent results for the operator $Q_1$ by a different group), and the FNAL/MILC \cite{Bazavov:2016nty} and HPQCD \cite{Dowdall:2019bea} lattice collaborations 
into updated predictions for the mass differences 
\begin{align}
\DeltaMd^\text{Average 2019} &= \left(0.533_{-0.036}^{+0.022}\right)\text{ps}^{-1} = \left(1.05_{-0.07}^{+0.04}\right) \DeltaMd^\text{exp} \, ,
\\
\DeltaMs^\text{Average 2019} &= \left(18.4_{-1.2}^{+0.7}\right)\text{ps}^{-1} = \left(1.04_{-0.07}^{+0.04}\right) \DeltaMs^\text{exp} \, .
\end{align}
These results are about 40\% more precise than the values in Eq.~\eqref{eq:DeltaMdFLAG2019} 
and Eq.~\eqref{eq:DeltaMsFLAG2019} obtained from the current FLAG~\cite{Aoki:2019cca} averages 
and show better agreement with the experimental results. The average for the SU(3)-breaking 
ratio $\xi$ that determines the SM predictions for $\DeltaMd/\DeltaMs$ is dominated 
by the SR calculation combined with the most recent lattice results for the decay constants, 
which provides a precision much better than the lattice-alone value for this quantity. 
The new averages agree very well with the individual SR results (see \cref{fig:MatrixElementsComparison}
and \cref{fig:XiComparison}), 
further reinforcing the usefulness of this as a totally independent determination of these 
important non-perturbative parameters. With our new average the hadronic uncertainties in 
these quantities are now at the same level as the CKM uncertainties. We then argued that with 
improved determinations of the hadronic matrix elements and the CKM elements the precision 
in $\DeltaMs$ can be improved to 3\% by about 2025. 

\begin{figure}
\includegraphics[width=0.47\textwidth]{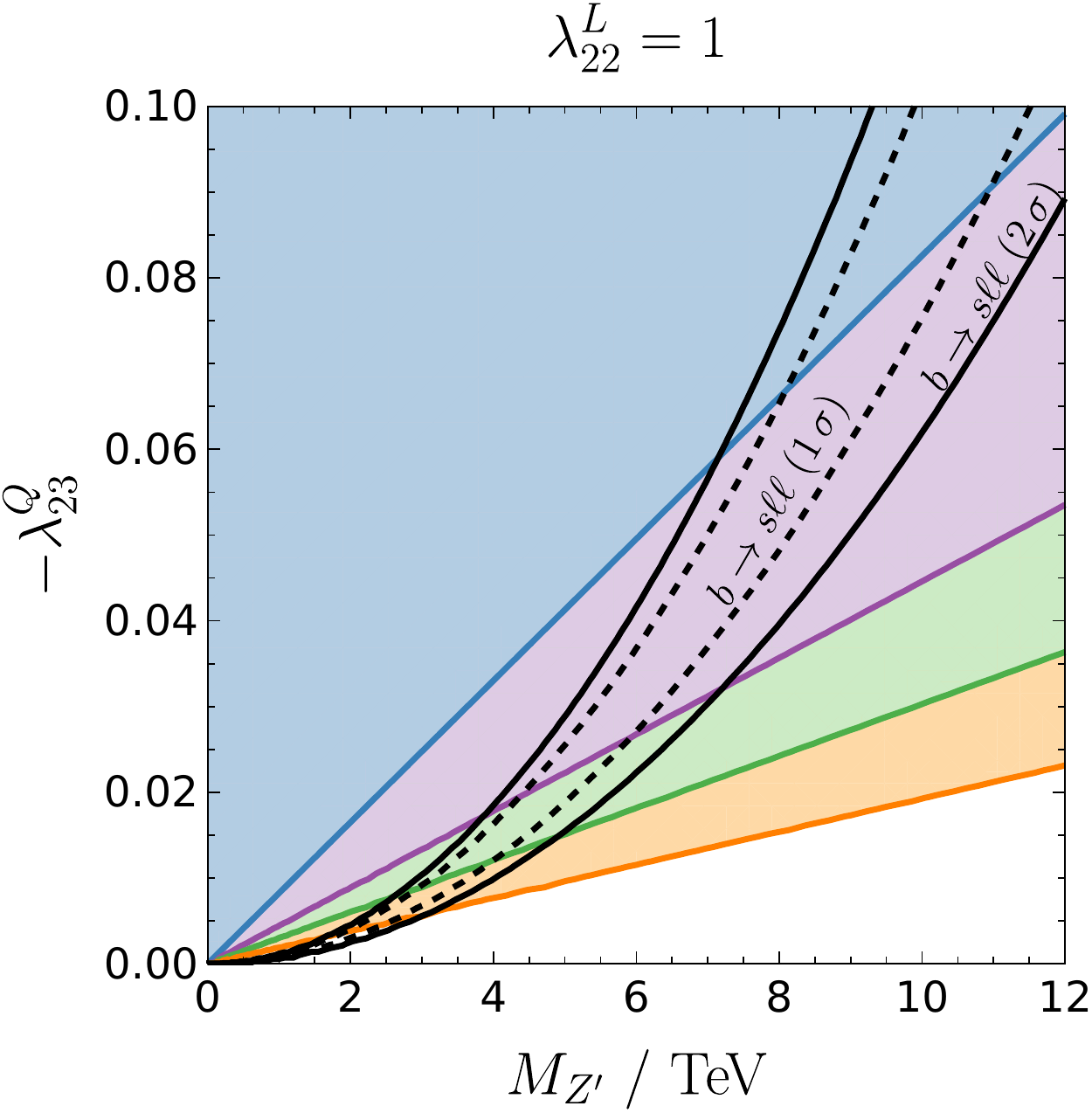}
\hfill
\includegraphics[width=0.47\textwidth]{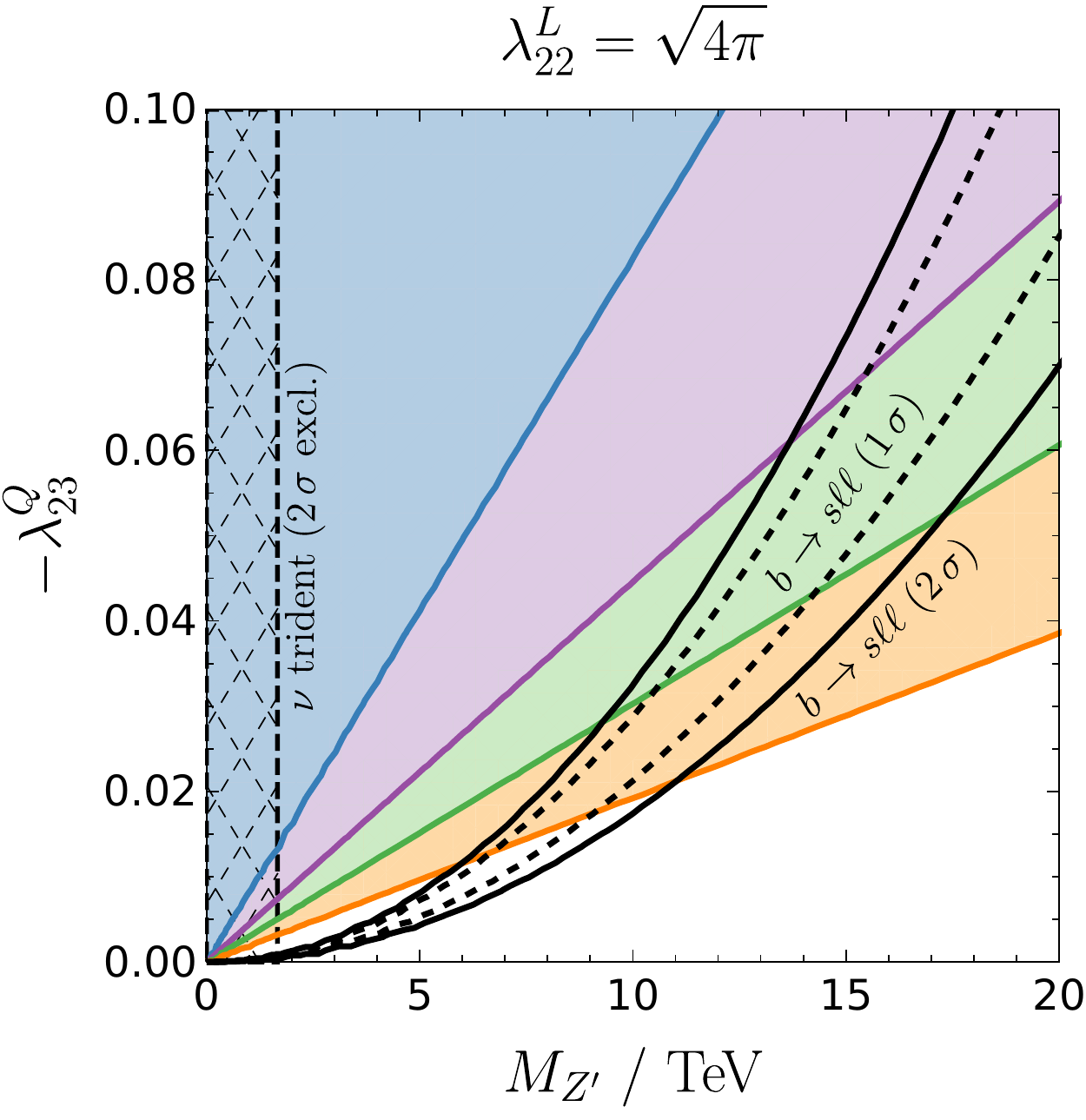}
\includegraphics[width=0.9\textwidth]{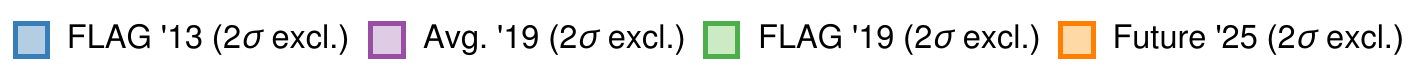}
\caption{Parameter space of a naive 
$Z'$ model explaining $b \to s\ell\ell $ via real LH currents, for two representative lepton 
couplings: $\lambda^L_{22}=1$ (left panel) and $\lambda^L_{22}=\sqrt{4\pi}$ (right panel). 
Constraints from various determinations of $\Delta M_s$ and neutrino trident production are 
shown as well.}
\label{fig:MixingvsRK_all}
\end{figure}

We investigated the constraints from $\DeltaMs$ on minimal $Z'$ and LQ explanations of 
the $b\to s\ell\ell$ anomalies. The constraints for LQs are generally weaker because 
contributions to $B_s$ mixing are only generated at loop level. Still we find that the 
assumed Future~'25 scenario leads to an upper bound of \SI{30}{\TeV} on the LQ mass, stronger than 
the unitarity constraint \cite{DiLuzio:2017chi,DiLuzio:2016sur} by more than a factor of 
two. The situation for the $Z'$ is summarized in \cref{fig:MixingvsRK_all}. Despite 
the reduction of the uncertainty the constraints with the updated prediction Average~'19 
are \emph{weaker} by about a factor $\sfrac{3}{2}$ than those obtained with the FLAG~'19 
scenario like in our earlier work \cite{DiLuzio:2017fdq} because of the smaller central 
value. Nevertheless, $B_s$ mixing by itself is sufficient to exclude the minimal $Z'$ 
scenario with a benchmark value of $\lambda^L_{22}=1$ for the lepton coupling for $Z'$ 
masses above \SI{6}{\TeV}. Assuming the Future~'25 scenario, we obtain the upper limit 
$M_{Z'}\lesssim \SI{9}{\TeV}$ for lepton couplings 
saturating the perturbativity bound (see right panel in \cref{fig:MixingvsRK_all}), which is about 
six times better than the unitarity bound on the $Z'$ mass. 
On the other hand, the left plot in \cref{fig:MixingvsRK_all} 
demonstrates that the parameter space will be very strongly constrained 
for more perturbative values of $\lambda^L_{22}$. 
An interesting open question is to what extent this mass window can be covered by 
direct searches at different future colliders. 

Last but not least, we have addressed the question whether an extension of the minimal 
$Z'$ scenario might relax the strong constraints. We have discussed two possibilities 
in which the minimal model might be extended, by adding new CP violating phases to the 
quark coupling, or including a coupling to both LH and RH quarks. For the case of new 
CP phases, the \PBs mixing observable \(\Acpmix\) is very strongly constraining unless 
the real part is sufficiently small, which prevents an explanation of the anomalies.
Adding instead a coupling to RH quarks has been recently discussed in the literature 
following the update, earlier this year at Moriond, of the LHCb measurement of \(\RK\). 
We demonstrate that the currently observed pattern, with \(\RKstar < \RK < 1\) requires 
a particular sign combination for the LH and RH quark couplings, which would lead to a 
positive shift to \(\DeltaMs\) rather than a reduction. 
We conclude that the constraints from $B_s$ mixing  
cannot be easily avoided within this class of minimally extended 
$Z'$ simplified models, although more general constructions can 
achieve that. 

All in all, we have shown how we are now entering the age of precision determinations 
of \(\DeltaMs\) and that, as we proceed further, the latter will become 
an increasingly powerful tool in order to constrain 
new physics explanations of other flavour observables.

\section*{Acknowledgements} 
We thank Matheus Hostert for useful discussions. 
The work of LDL was supported by the ERC grant NEO-NAT. The work of AL is supported by the STFC grant of the IPPP.
The work of MK was supported by MIUR (Italy) under a contract PRIN 2015P5SBHT and by INFN Sezione di Roma La Sapienza and partially supported by the ERC-2010 DaMESyFla Grant Agreement Number: 267985.
The authors thank Ben Allanach for noticing some missing text in the first version of this paper.

\appendix

\renewcommand\thetable{\arabic{table}} 

\section{Input parameters and average matrix elements\label{sec:inputs}}

We list the input parameters required for the evaluation of the mass differences 
in Table~\ref{tab:other_inputs}. 
We note that the CKM parameters are taken from the standard 
fit of the CKMfitter collaboration \cite{Charles:2004jd,CKMfitter:Summer18} which includes 
$\DeltaMs$ and $\DeltaMd$ as constraints. The FNAL/MILC~\cite{Bazavov:2016nty} and 
HPQCD~\cite{Dowdall:2019bea} collaborations instead use the result of CKMfitter's 
tree fit
\begin{equation}
 \text{tree fit: } \V{cb} = (42.41^{+0.40}_{-1.51})\,\times 10{-3}\,,
\end{equation}
which uses only tree-level observables and is therefore independent of the mass differences. 
However, also a number of other observables are discarded in this approach. Using CKMlive \cite{CKMlive} 
we have performed a fit where only $\DeltaMs$ and $\DeltaMd$ were excluded which yields 
\begin{equation}
\text{fit without } \Delta M_{s,d} \text{: } \V{cb} = (42.40^{+0.40}_{-1.17})\,\times 10{-3}\,.
\end{equation}
We observe that this result is very close to that from the standard fit shown in 
Table~\ref{tab:other_inputs} and therefore use the standard fit for simplicity.
\begin{table}
\begin{tabular}{@{}LLl@{}}
\toprule
\text{Parameter} & \text{Value} & Source \\
\midrule
M_Z & \SI{91.1876(21)}{\GeV} & PDG 2019 \cite{Tanabashi:2018oca,PDG:online} \\
\alpha_s (M_Z) & \num{0.1181(11)} & PDG 2019 \cite{Tanabashi:2018oca,PDG:online} \\
m_{t, \text{pole}} & \SI{173.1 \pm 0.9}{\GeV} & PDG 2019 \cite{Tanabashi:2018oca,PDG:online} \\
\overline{m}_b(\overline{m}_b) & \SI[parse-numbers=false]{(4.18^{+0.03}_{-0.02})}{\GeV} & PDG 2019 \cite{Tanabashi:2018oca,PDG:online} \\
\addlinespace
\V{us} & 0.224745^{+0.000254}_{-0.000059} & CKMfitter Summer 2018 \cite{Charles:2004jd,CKMfitter:Summer18} \\
\V{ub} & 0.003746^{+0.000090}_{-0.000062} & CKMfitter Summer 2018 \cite{Charles:2004jd,CKMfitter:Summer18} \\
\V{cb} & 0.04240^{+0.00030}_{-0.00115} & CKMfitter Summer 2018 \cite{Charles:2004jd,CKMfitter:Summer18} \\
\gamma_\text{CKM} & \SI[parse-numbers=false]{(65.81^{+0.99}_{-1.61})}{\degree} & CKMfitter Summer 2018 \cite{Charles:2004jd,CKMfitter:Summer18} \\
\addlinespace
\overline{m}_t(\overline{m}_t) & \SI{163.3 \pm 0.9}{\GeV} & Own evaluation (RunDec \cite{Chetyrkin:2000yt,Herren:2017osy}) \\
\addlinespace
f_{B_s} & \SI{230.3 \pm 1.3}{\MeV} & FLAG 2019 \cite{Aoki:2019cca}\\
f_{B} & \SI{190.0 \pm 1.3}{\MeV} & FLAG 2019 \cite{Aoki:2019cca}\\
\bottomrule
\end{tabular}
\caption{
Input parameters for our calculations of $\DeltaMs$ and $\DeltaMd$.
}
\label{tab:other_inputs}
\end{table}

In addition to our averages given in \cref{eq:BsAverageMatrixElements} we also provide the weighted averages for the 
matrix elements in the $B_d$ system 
\begin{align}
 f_{B}^2B_1^d(\mu_b) & = (0.0305\pm0.0011)\,\text{GeV}^2\,,\nonumber\\
 f_{B}^2B_2^d(\mu_b) & = (0.0288\pm0.0013)\,\text{GeV}^2\,,\nonumber\\
 f_{B}^2B_3^d(\mu_b) & = (0.0281\pm0.0020)\,\text{GeV}^2\,,\nonumber\\
 f_{B}^2B_4^d(\mu_b) & = (0.0387\pm0.0015)\,\text{GeV}^2\,,\nonumber\\
 f_{B}^2B_5^d(\mu_b) & = (0.0361\pm0.0014)\,\text{GeV}^2\,.
 \label{eq:BdAverageMatrixElements}
\end{align}
and the weighted averages for the bag parameters 
\begin{equation}
 \begin{array}{ll}
  B_1^s(\mu_b) = 0.849\pm0.023\,, &\hspace{1cm}B_1^d(\mu_b) = 0.835\pm0.028\,,\nonumber\\
  B_2^s(\mu_b) = 0.835\pm0.032\,, &\hspace{1cm}B_2^d(\mu_b) = 0.791\pm0.034\,,\nonumber\\
  B_3^s(\mu_b) = 0.854\pm0.051\,, &\hspace{1cm}B_3^d(\mu_b) = 0.775\pm0.054\,,\nonumber\\
  B_4^s(\mu_b) = 1.031\pm0.035\,, &\hspace{1cm}B_4^d(\mu_b) = 1.063\pm0.041\,,\nonumber\\
  B_5^s(\mu_b) = 0.959\pm0.031\,, &\hspace{1cm}B_5^d(\mu_b) = 0.994\pm0.037\,,
 \end{array}
 \label{eq:AberageBagParameters}
\end{equation}
at the scale $\mu_b=\bar{m}_b(\bar{m}_b)$.

\bibliographystyle{JHEP.bst}
\bibliography{bibliography}

\end{document}